# A Study of Primordial Very Massive Star Evolution. II. Stellar Rotation and Gamma-Ray Burst Progenitors

Guglielmo Volpato[1,2], Paola Marigo[1], Guglielmo Costa[1,2,3,4], Alessandro Bressan[5], Michele Trabucchi[1], Léo Girardi[2], and Francesco Addari[5]

[1] Dipartimento di Fisica e Astronomia Galileo Galilei, Università degli studi di Padova, Vicolo dell'Osservatorio 3, I-35122 Padova, Italy; guglielmo.volpato@phd.unipd.it
[2] Osservatorio Astronomico di Padova-INAF, Vicolo dell'Osservatorio 5, I-35122 Padova, Italy
[3] INFN-Padova, via Marzolo 8, I-35131 Padova, Italy
[4] Univ Lyon, Univ Lyon1, Ens de Lyon, CNRS, Centre de Recherche Astrophysique de Lyon UMR5574, F-69230 Saint-Genis-Laval, France
[5] SISSA, via Bonomea 365, I-34136 Trieste, Italy

## Abstract

We calculate new evolutionary models of rotating primordial very massive stars, with initial mass from $100\,M_\odot$ to $200\,M_\odot$, for two values of the initial metallicity $Z = 0$ and $Z = 0.0002$. For the first time in this mass range, we consider stellar rotation and pulsation-driven mass loss, along with radiative winds. The models evolve from the zero-age main sequence until the onset of pair-instability. We discuss the main properties of the models during their evolution and then focus on the final fate and the possible progenitors of jet-driven events. All tracks that undergo pulsational-pair instability produce successful gamma-ray bursts (GRB) in the collapsar framework, while those that collapse directly to black holes (BH) produce jet-driven supernova events. In these latter cases, the expected black hole mass changes due to the jet propagation inside the progenitor, resulting in different models that should produce BH within the pair-instability black hole mass gap. Successful GRBs predicted here from zero metallicity, and very metal-poor progenitors, may be bright enough to be detected even up to redshift ∼20 using current telescopes such as the Swift-BAT X-ray detector and the JWST.

*Unified Astronomy Thesaurus concepts:* Stellar evolution (1599); Stellar remnants (1627); Stellar rotation (1629); Population III stars (1285); Stellar winds (1636); Gamma-ray bursts (629)

## 1. Introduction

The first stars (Population III) that lit up in our Universe might have been far more massive than those forming nowadays, with an initial mass function peaking at $\simeq 100\,M_\odot$ (Bromm et al. 1999; Abel et al. 2002). The reason is the absence of metals, which are the most efficient coolants within molecular clouds during star formation. In a previous work (Volpato et al. 2023, hereafter, Paper I, to which we refer for an introduction on Population III and very metal-poor stars), we investigated the effect of pulsation-driven mass loss (see the prescription in Nakauchi et al. 2020) on the evolution and final fate of primordial very massive stars (VMSs). We found that pulsation-driven and radiation-driven mass loss dominates during different evolutionary phases. In Paper I, the most massive stars can eject several solar masses of material already during the main sequence (MS) due to radial pulsations. This is in contrast with the modest mass loss expected from radiation-driven winds at these metallicities. We find that almost all models experience pair-creation instability during the last phases of the evolution after the ignition of carbon, neon, or oxygen, depending on the mass of the star. Models with $M_i = 100\,M_\odot$ and $Z = 0$ are somewhat uncertain and may avoid pair-instability. Our stars with $300 \leqslant M_i/M_\odot \leqslant 1000$ should directly collapse into a black hole (DBH, Bond et al. 1984; Farmer et al. 2020), while models with $M_i = 150, 200\,M_\odot$ should produce pair-instability supernovae (PISN), leaving no remnant (Fowler & Hoyle 1964; Barkat et al. 1967; Rakavy & Shaviv 1967; Fraley 1968; Heger & Woosley 2002; Heger et al. 2003; Takahashi et al. 2018; Takahashi 2018; Woosley & Heger 2021; Farag et al. 2022; Costa et al. 2023). Models with $M_i = 100\,M_\odot$ and $Z = 0.0002$ should enter into the pulsational pair-instability supernova (PPISN) regime (Woosley et al. 2002; Chen et al. 2014; Yoshida et al. 2016; Woosley 2017; Farmer et al. 2019; Woosley & Heger 2021; Farag et al. 2022). Instead, those with the same mass and $Z = 0$ could end their evolution either with a failed core-collapse supernova (CCSN) or a PPISN. In the former case, these models could provide a new formation pathway for black holes (BHs), potentially helping to alleviate the black hole mass gap puzzle (see also Croon et al. 2020; Farmer et al. 2020; Sakstein et al. 2020; Costa et al. 2021; Farrell et al. 2021; Tanikawa et al. 2021; Vink et al. 2021; Farag et al. 2022, for different formation scenarios).

In Paper I, we did not consider stellar rotation, which is one of the most influential phenomena in the evolution of massive and very massive stars (Heger et al. 2000; Meynet & Maeder 2000; Brott et al. 2011; Ekström et al. 2012; Paxton et al. 2013; Yusof et al. 2013; Limongi & Chieffi 2018; Goswami et al. 2021, 2022; Higgins et al. 2022; Martinet et al. 2023). Rotation changes the gravity of the star and thus modifies the stellar geometry from the usual spherical symmetry also affecting the surface temperature (e.g., Costa et al. 2019a, 2019b, and references therein). The fact that the star is rotating implies different kinds of turbulent mixing, e.g., meridional circulation and shear instability. These processes increase the mixing of chemical elements within the star, affecting the lifetimes of the main nuclear burning phases as well as the chemical composition at the star's surface (Maeder 2009, and references therein). Turbulence caused by rotation transports angular momentum from the stellar core to the envelope (Heger et al. 2000), where it is removed by stellar winds. Different authors (e.g., Heger et al. 2000; Georgy et al. 2011, and references therein) have suggested that mass loss is enhanced by rotation due to lower effective gravity, which is caused by centrifugal forces. However, the interplay between rotation and mass loss is still under debate, with different results reported in the literature. For instance, more sophisticated multidimensional radiation-driven wind models have not confirmed the enhancement of mass loss due to rotation (Müller & Vink 2014).

While the effect of rotation on mass loss in VMSs has been the subject of several studies (e.g., Meynet & Maeder 2000; Ekström et al. 2012; Yoon et al. 2012, 2015; Murphy et al. 2021; Martinet et al. 2023), its impact in combination with pulsation-induced mass loss has never been examined. In light of the results of Paper I, this topic clearly deserves to be investigated. We do so using the PARSEC code (Bressan et al. 2012; Costa et al. 2019a, 2019b; Nguyen et al. 2022) to follow the evolution of a set of VMSs until the occurrence of pair-instability, in line with our previous work.

Rotation can also heavily affect the final fate of massive and very massive stars, and it is a necessary condition to produce successful GRB events (Woosley 1993; Woosley & Heger 2006; Yoon et al. 2006, 2012, 2015; Woosley & Heger 2012). The collapsar model (Woosley 1993) is the most widely accepted theory for the formation of long GRBs, in which an accretion disk forms during the collapse of a massive star, powering the jets that produce the GRB event. The accretion disk can form only if the infalling material has enough angular momentum to avoid direct accretion onto the BH (Woosley 1993). The propagation of the jet through the progenitor is a major issue for massive stars with extended envelopes, preventing the production of a successful gamma-ray burst (GRB) event. The reason is that the crossing timescale for the jet to reach the surface of the star can be longer than the accretion timescale by more than 3 orders of magnitude. In this case, the jet is not powerful enough to break out from the star and produce a successful GRB (Yoon et al. 2006, 2012, 2015; Woosley & Heger 2012). This is why, when considering Population III stars, Yoon et al. (2012) proposed the chemically homogeneous evolution as the main channel for the star to retain enough angular momentum within its core and at the same time to avoid the redward evolution in the H-R diagram. In this way, the star does not retain a very extended envelope and this facilitates the jet propagation. In this work, we investigate models of VMS that should experience pair-instability (PI) but with lower masses compared to Yoon et al. (2015) to have less extended envelopes. We find another possible pathway for the evolution of successful GRB progenitors. The absence of the stellar envelope due to pulsational pair-instability mass loss eases the jet propagation through the star. This decreases enormously the crossing timescale for the jet to reach the stellar surface, thus producing successful GRB events.

The paper is organized as follows. Section 2 summarizes the main ingredients of the PARSEC code. In Section 3 we present stellar models computed with and without rotation accounting only for radiation-driven winds and for both radiation and pulsation-driven mass loss. We discuss the main aspects of the models, especially how dredge-up episodes affect the internal structure, the surface abundances, and the wind ejecta. Then, we focus on the final fates, the compact remnants, and the possible jet-driven events for different models' structures and angular momentum configurations. In Section 4 we summarize our results and conclude.

## 2. Stellar Evolution Calculations

We compute stellar evolution calculations with the PARSEC code (Bressan et al. 2012) in its version 2.0 (Costa et al. 2019a, 2019b, 2021; Nguyen et al. 2022, and reference therein). The input physics and the code parameters are as described in Paper I, whereas the details of the implementation of rotation in the PARSEC code can be found in Costa et al. (2019a, 2019b). We use scaled-solar abundances (Caffau et al. 2011), where the initial solar metallicity follows the calibration in Bressan et al. (2012), $Z_{\mathrm{initial},\odot} = 0.01774$. Here, we briefly summarize the main aspects of the PARSEC code that concern the evolution of massive stars and, in particular, the mass loss prescriptions adopted in this work. For the standard mass loss prescription ($\dot{M}_{\mathrm{rdw}}$ for consistency with Paper I), we adopt the formulation by Vink et al. (2000, 2001) and the mass loss rates predicted by de Jager et al. (1988) for stars with an effective temperature higher and lower than 10,000 K, respectively. The metallicity dependence is $\dot{M} \propto (Z/Z_{\mathrm{initial},\odot})^{0.85}$, where we use the initial metallicity as a proxy for the iron content[6] (which remains constant along the evolution). We also consider the enhancement of mass loss due to the proximity of the Eddington factor to 1 (Gräfener & Hamann 2008; Vink et al. 2011). In this case, we use the same scheme as in Chen et al. (2015), with the following details $\dot{M} \propto (Z/Z_{\mathrm{initial},\odot})^{\alpha}$, where $\alpha = 2.45 - 2.4\Gamma_{\mathrm{e}}$, with $\Gamma_{\mathrm{e}}$ the Eddington factor and $\alpha$ between 0 and 0.85. In each evolutionary stage, we take the maximum mass loss rate between Vink et al. (2000, 2001) or de Jager et al. (1988) and Vink et al. (2011), which accounts for the Eddington factor dependence. For Wolf–Rayet stars with $X < 0.3$ and $\log(T_{\mathrm{eff}}) > 4$, we use the mass loss recipe from Sander et al. (2019) with the metallicity dependence proposed by Costa et al. (2021).

For pulsation-driven mass loss ($\dot{M}_{\mathrm{pdw}}$), we use the analytical expressions provided by Nakauchi et al. (2020)

$$\log\left(\frac{\dot{M}_{\mathrm{pdw}}}{M_\odot\,\mathrm{yr}^{-1}}\right) = \alpha_1 \log\left(\frac{M}{10^3\,M_\odot}\right) - \alpha_2 \\ - \beta_1[\log(T_{\mathrm{eff}}) - \beta_2]^\gamma \quad (1)$$

$$\log\left(\frac{\dot{M}_{\mathrm{pdw}}}{M_\odot\,\mathrm{yr}^{-1}}\right) = -2.88 + \log\left(\frac{M}{10^3\,M_\odot}\right) \\ - 15.6[\log(T_{\mathrm{eff}}) - 3.7], \quad (2)$$

where $\alpha_1$, $\alpha_2$, $\beta_1$, $\beta_2$, and $\gamma$ are coefficients that depend on the initial metallicity of the model (see Nakauchi et al. 2020, for more details). Finally, the last mass loss prescription is $\dot{M}_{\mathrm{max}}$, which takes the maximum between $\dot{M}_{\mathrm{rdw}}$ and $\dot{M}_{\mathrm{pdw}}$ at each time step during the evolution of the models.

Concerning the interplay between mass loss and stellar rotation, we use the prescription by Heger et al. (2000), which

---

[6] when using Vink et al. (2000, 2001) prescription, the initial stellar iron content has to be rescaled to the initial solar value according to our calibration

reads

$$\dot{M}(\omega) = \dot{M}(\omega = 0)\left(\frac{1}{1 - v/v_{\text{crit}}}\right)^{\xi}, \text{ with } \xi = 0.43, \quad (3)$$

where $\dot{M}(\omega = 0)$ is the mass loss rate in the nonrotating case. Then, $v$ is the surface tangential stellar velocity, while $v_{\text{crit}}$ is the break-up velocity defined as

$$v_{\text{crit}}^2 = \frac{Gm}{r}(1 - \Gamma_e). \quad (4)$$

We take into account also the mechanical mass loss when the star reaches the critical rotation (Georgy et al. 2013). It is worth noticing that there are different prescriptions for the treatment of stellar rotation (e.g., Maeder & Meynet 2000), and this subject is still under investigation. Major uncertainties in stellar rotation are tied to different aspects. For instance, stellar rotation is inextricably linked to magnetic field generation (e.g., Braithwaite & Spruit 2017; Brun & Browning 2017). The interaction between rotation and magnetic fields is not completely understood despite its influence on stellar activity, which affects processes such as starspots, flares, and the stellar wind. Furthermore, rotation influences mass and angular momentum loss from stars, thus affecting their evolution. Predicting the rate and pattern of mass loss due to rotation-induced instabilities is still a source of uncertainties in evolutionary models.

We adopt the mixing-length theory by Böhm-Vitense (1958), with a mixing-length parameter $\alpha_{\text{mlt}} = 1.74$. We use the Schwarzschild criterion to define the border of the convective regions, while for core overshooting, we adopt the ballistic approximation (Bressan et al. 1981). In this latter, the core overshooting parameter ($\lambda_{\text{ov}} = 0.4$) times the pressure scale height corresponds to the mean free path of the eddies *across* the border of the convective region. We also account for overshooting at the base of the convective envelope below the formal Schwarzschild border (Alongi et al. 1991; Bressan et al. 2012; Nguyen et al. 2022). Recent calibration of the red giant branch bump luminosity in a large sample of Globular Clusters with metallicity $-2 < [M/H] < 0$ using updated $\alpha$-enhanced PARSEC models (Fu et al. 2018) suggests using an envelope overshooting parameter $\Lambda_{\text{env}}$ between 0.5 and 0.7, with the latter value being more appropriate in more metal-poor systems. Similar, if not greater, values are required to reproduce blue loops in star clusters and in low metallicity dwarf irregular galaxies (Alongi et al. 1991; Bressan et al. 2012; Tang et al. 2014). In the present calculations, we use $\Lambda_{\text{env}} = 0.7$.

To inhibit density inversion in the inefficient convective regions of the stellar envelope, we follow the temperature gradient limitation described in Chen et al. (2015). By imposing $\nabla_T \leqslant \nabla_{T_{\text{max}}} = \frac{1 - \chi_\mu \nabla_\mu}{\chi_T}$, convection becomes more efficient, preventing the numerical instabilities caused by density inversion.

As in Paper I, we consider two different initial chemical compositions ($Z = 0$, $Y = 0.2485$) and ($Z = 0.0002$, $Y = 0.24885$), with $Z$ and $Y$ the initial abundances in mass fraction of metals and helium, respectively. We compute stellar evolution models with initial mass $M_i = 100, 150, 200 \, M_\odot$ and *initial* rotation rate $\omega = 0.0, 0.2, 0.3, 0.4, 0.5$; where $\omega = \Omega_i/\Omega_{\text{crit}}$ with $\Omega_i$ the initial angular velocity and $\Omega_{\text{crit}}$ the critical angular velocity. For each combination of ($M_i$, $Z$, $\omega$), we compute two stellar models adopting two different mass loss recipes, namely $\dot{M}_{\text{rdw}}$ and $\dot{M}_{\text{max}}$. We found that models encounter progressively greater numerical difficulties in the computation toward the highest values of mass, rotation velocity, and metallicity in the explored range. In particular, two models out of 60 could not be brought to convergence (the models with $M_i = 150 \, M_\odot$ and $M_i = 200 \, M_\odot$, having $Z = 0.0002$, $\omega_i = 0.5$, and computed with the $\dot{M}_{\text{rdw}}$ prescription), so we exclude them from the following discussion.

We note that the calculations of Nakauchi et al. (2020) are based on nonrotating models, while in the present study, the interplay between rotation and stellar pulsation should be taken into account. However, we found the timescale of pulsation to be always much shorter than the rotation period in all our models, indicating that the interaction between the two processes can be safely neglected (see Appendix C for further details).

We follow the evolution of our models from the zero-age main sequence (ZAMS) until the onset of pair-creation instability (see Section 2 in Paper I). This occurs after the ignition of carbon, neon, and oxygen in the stellar core, depending on the initial mass, metallicity, rotation, and mass-loss prescription adopted for the models (see Tables 1 and 2).

## 3. Results

### 3.1. General Properties of the Stellar Evolutionary Tracks

In Table 1 and 2 we summarize the main properties that highlight the evolution and final outcome of the models.

Figure 1 presents the evolution of all models in the H-R diagram (HRD), where we can see that there is a positive correlation between the models' luminosity and their initial rotation velocity, $\omega$. This is most evident for models computed with $Z = 0.0002$, while in the case of $Z = 0$ the evolutionary tracks run almost superimposed except for the nonrotating ones. This behavior can be explained by the helium enrichment at the surface during MS, which forces both luminosity and effective temperature to increase. In turn, this helium enrichment is caused by two factors. First, the enhancement of the convective core between rotating and nonrotating models increases with metallicity (Groh et al. 2019). Second, at $Z = 0.0002$, the region above the convective core experiences greater rotational mixing as the rotation rate increases, compared to the $Z = 0$ models at the same rotation rates.

Another effect of rotation is to reduce or even quench the blue loops during the core helium burning (cHeB) phase. We can see that in all panels of Figure 1, there is at least one nonrotating star that evolves toward higher effective temperatures. With the addition of rotation, this is not the case anymore. Models with $Z = 0$ and $\omega > 0$ evolve after the MS toward the red part of the HRD; while for $Z = 0.0002$ there are two models with $\omega = 0.2$ that perform a blue loop, but in the case of $\omega > 0.2$ no model evolves back to higher effective temperatures.

After the MS phase, due to rotational mixing and the occurrence of dredge-up (DUP) episodes, 16 stellar models reach a surface hydrogen abundance $X < 0.3$ (starred symbols in Figure 1). When massive stars evolve into red supergiants, the convective envelope inflates and cools, while at low densities, opacity is dominated by electron scattering. These lead to increasing atmospheric opacity and favor the development of convection in progressively deeper layers of the star, causing a DUP episode. It is worth noticing that the efficiency

Table 1
Most Relevant Properties of Models Computed with $Z = 0.0002$, $\dot{M}_{\rm rdw}$ and $\dot{M}_{\rm max}$

| $M_i$ ($M_\odot$) (1) | $\tau_{\rm MS}$ (Myr) (2) | $\tau_{\rm cHeB}$ (Myr) (3) | $f_{\rm H\,puls}$ (4) | $f_{\rm He\,puls}$ (5) | Blue Loop (6) | DUP (7) | $M_{\rm He}$ ($M_\odot$) (8) | $M_{\rm CO}$ ($M_\odot$) (9) | $M_f$ ($M_\odot$) (10) | $X_{\rm core}$ Onset PI (11) | $L_\nu/L_{\rm rad}$ ($\log_{10}$) (12) | Fate (13) | Remnant (14) | $M_{\rm BH}$ ($M_\odot$) (15) |
|---|---|---|---|---|---|---|---|---|---|---|---|---|---|---|
| | | | | | | | $\dot{M}_{\rm rdw}$ $\omega = 0.0$ | | | | | | | |
| 100 | 2.83 | 0.25 | 0.29 | 0.57 | × | ✓ | 53.8 | 47.6 | 94.3 | 0.865 O | 3.1 | PPISN | BH | 40.9 |
| 150 | 2.45 | 0.24 | 0.45 | 0.32 | ✓ | ✓ | 79.5 | 71.8 | 146.4 | 0.080 Ne | 3.2 | PISN | × | ... |
| 200 | 2.25 | 0.23 | 0.50 | 0.33 | ✓ | ✓ | 110.3 | 100.7 | 193.9 | 0.003 C | 3.2 | PISN | × | ... |
| | | | | | | | $\omega = 0.2$ | | | | | | | |
| 100 | 2.87 | 0.26 | 0.28 | 0.96 | × | × | 52.0 | 46.0 | 93.4 | 0.853 O | 3.1 | PPISN | BH | 39.8 |
| 150 | 2.48 | 0.25 | 0.45 | 0.97 | × | ✓ | 45.1 | 38.8 | 143.0 | 0.131 O | 2.4 | PPISN | BH | 40.09 |
| | | | | | | | | | | | | DBH[b] | BH | 128.7 |
| 200 | 2.28 | 0.24 | 0.49 | 0.76 | ✓ | ✓ | 51.9 | 46.0 | 191.5 | 0.090 O | 2.4 | PPISN | BH | 48.5 |
| | | | | | | | | | | | | DBH[b] | BH | 172.4 |
| | | | | | | | $\omega = 0.3$ | | | | | | | |
| 100 | 2.92 | 0.27 | 0.27 | 0.97 | × | × | 51.7 | 45.3 | 93.1 | 0.848 O | 3.1 | PPISN | BH | 39.5 |
| 150 | 2.51 | 0.25 | 0.46 | 0.97 | × | × | 80.9 | 71.8 | 142.8 | 0.061 Ne | 3.2 | PISN | × | ... |
| 200 | 2.30 | 0.23 | 0.51 | 0.98 | × | × | 108.9 | 98.1 | 191.1 | 0.001 C | 3.2 | PISN | × | ... |
| | | | | | | | $\omega = 0.4$ | | | | | | | |
| 100 | 2.98 | 0.27 | 0.26 | 0.95 | × | ✓ | 51.5 | 44.5 | 92.9 | 0.877 O | 3.1 | PPISN | BH | 39.4 |
| 150 | 2.56 | 0.24 | 0.50 | 0.98 | × | ✓ | 77.6 | 69.6 | 142.3 | 0.020 Ne | 3.1 | PISN | × | ... |
| 200 | 2.33 | 0.23 | 0.53 | 0.90 | × | ✓ | 95.2 | 95.0 | 189.4 | 0.022 Ne | 2.7 | PISN | × | ... |
| | | | | | | | | | | | | DBH[b] | BH | 170.5 |
| | | | | | | | $\omega = 0.5$ | | | | | | | |
| 100 | 3.09 | 0.27 | 0.22 | 0.53 | × | ✓ | 53.2 | 45.8 | 95.0 | 0.879 O | 3.1 | PPISN | BH | 40.7 |
| | | | | | | | | | | | | PISN[a] | × | ... |
| | | | | | | | $\dot{M}_{\rm max}$ $\omega = 0.0$ | | | | | | | |
| 100 | 2.84 | 0.26 | 0.28 | 0.95 | × | ✓ | 53.1 | 46.9 | 92.7 | 0.859 O | 3.1 | PPISN | BH | 40.4 |
| 150 | 2.47 | 0.24 | 0.44 | 0.38 | ✓ | ✓ | 77.0 | 69.3 | 139.7 | 0.055 Ne | 3.2 | PISN | × | ... |
| 200 | 2.29 | 0.23 | 0.49 | 0.36 | ✓ | ✓ | 105.7 | 96.2 | 180.2 | 0.002 C | 3.2 | PISN | × | ... |
| | | | | | | | $\omega = 0.2$ | | | | | | | |
| 100 | 2.88 | 0.27 | 0.28 | 0.97 | × | × | 52.0 | 46.0 | 92.2 | 0.857 O | 3.1 | PPISN | BH | 39.6 |
| 150 | 2.50 | 0.24 | 0.46 | 0.98 | × | × | 81.0 | 74.1 | 136.7 | 0.086 Ne | 3.3 | PISN | × | ... |
| 200 | 2.31 | 0.23 | 0.52 | 0.49 | ✓ | ✓ | 91.2 | 91.2 | 176.6 | 0.021 Ne | 2.8 | PISN | × | ... |
| | | | | | | | | | | | | DBH[b] | BH | 158.9 |
| | | | | | | | $\omega = 0.3$ | | | | | | | |
| 100 | 2.91 | 0.27 | 0.25 | 0.97 | × | × | 49.6 | 43.6 | 92.1 | 0.768 O | 3.1 | PPISN | BH | 38.2 |
| 150 | 2.53 | 0.24 | 0.50 | 0.99 | × | ✓ | 71.4 | 67.6 | 135.7 | 0.883 O | 2.8 | PISN | × | ... |
| | | | | | | | | | | | | DBH[b] | BH | 122.1 |
| 200 | 2.34 | 0.23 | 0.58 | 0.98 | × | ✓ | 101.1 | 99.1 | 174.3 | 0.117 Ne | 2.9 | PISN | × | ... |
| | | | | | | | | | | | | DBH[b] | BH | 156.9 |
| | | | | | | | $\omega = 0.4$ | | | | | | | |
| 100 | 2.99 | 0.27 | 0.28 | 0.97 | × | × | 51.1 | 44.4 | 91.8 | 0.880 O | 3.1 | PPISN | BH | 39.1 |
| 150 | 2.58 | 0.25 | 0.55 | 0.98 | × | ✓ | 77.1 | 69.0 | 134.6 | 0.009 Ne | 3.1 | PISN | × | ... |
| | | | | | | | | | | | | DBH[a,b] | BH | 121.1 |
| 200 | 2.37 | 0.24 | 0.65 | 0.98 | × | ✓ | 93.8 | 89.5 | 171.3 | 0.101 Ne | 2.9 | PISN | × | ... |
| | | | | | | | | | | | | DBH[b] | BH | 154.2 |
| | | | | | | | $\omega = 0.5$ | | | | | | | |
| 100 | 3.06 | 0.27 | 0.06 | 0.94 | × | × | 54.0 | 46.5 | 91.7 | 0.878 O | 3.1 | PPISN | BH | 40.8 |
| 150 | 2.62 | 0.25 | 0.63 | 0.96 | × | ✓ | 81.7 | 71.5 | 132.2 | 0.029 Ne | 3.2 | PISN | × | ... |
| 200 | 2.41 | 0.24 | 0.75 | 0.95 | × | × | 114.6 | 101.7 | 163.1 | 0.001 C | 3.3 | PISN | × | ... |

**Notes.** The table entries are as follows: (1) star's initial mass; (2) MS lifetime; (3) cHeB lifetime; (4) and (5) fractions of MS and cHeB lifetimes in which the star is unstable to radial pulsation; (6) and (7) occurrence of blue loop and dredge-up episode; (8) final He core mass; (9) final C-O core mass; (10) final mass of the star at the onset of dynamical instability; (11) central fuel abundance of ongoing nuclear burning at the onset of dynamical instability; (12) neutrino luminosity to radiative luminosity ratio when $T_c = 10^9$ K; (13) and (14) final fate and associated outcome (BH or complete disruption), and (15) BH mass.
[a] Assuming an error of 1% on the upper limit for PISN in the fit formula by Mapelli et al. (2020).
[b] Considering $M_{\rm He} = M_f$, so $M_{\rm BH} = 0.9 \cdot M_f$.

Table 2
Most Relevant Properties of Models Computed with $Z = 0$, $\dot{M}_{rdw}$ and $\dot{M}_{max}$

| $M_i$ ($M_\odot$) (1) | $\tau_{MS}$ (Myr) (2) | $\tau_{cHeB}$ (Myr) (3) | $f_{H\,puls}$ (4) | $f_{He\,puls}$ (5) | Blue Loop (6) | DUP (7) | $M_{He}$ ($M_\odot$) (8) | $M_{CO}$ ($M_\odot$) (9) | $M_f$ ($M_\odot$) (10) | $X_{core}$ Onset PI (11) | $L_\nu/L_{rad}$ ($\log_{10}$) (12) | Fate (13) | Remnant (14) | $M_{BH}$ ($M_\odot$) (15) |
|---|---|---|---|---|---|---|---|---|---|---|---|---|---|---|
| | | | | | $\dot{M}_{rdw}$ $\omega = 0.0$ | | | | | | | | | |
| 100 | 2.54 | 0.25 | 0.07 | 0.51 | ✓ | ✓ | 41.8 | 38.4 | 99.9 | 0.511 O | 2.7 | fCCSN[a] | BH | 89.9 |
| | | | | | | | | | | | | PPISN[b] | BH | 34.2 |
| 150 | 2.33 | 0.23 | 0.30 | 0.36 | ✓ | ✓ | 74.4 | 67.7 | 149.9 | 0.011 Ne | 3.1 | PISN | × | ⋯ |
| 200 | 2.16 | 0.22 | 0.27 | 0.35 | ✓ | ✓ | 110.4 | 103.8 | 199.9 | 0.001 C | 3.2 | PISN | × | ⋯ |
| | | | | | $\omega = 0.2$ | | | | | | | | | |
| 100 | 2.65 | 0.25 | 0.46 | 0.11 | × | ✓ | 50.6 | 45.4 | 99.9 | 0.858 O | 3.2 | PPISN | BH | 39.5 |
| 150 | 2.30 | 0.23 | 0.51 | 0.11 | × | × | 78.7 | 70.6 | 149.9 | 0.049 Ne | 3.2 | PISN | × | ⋯ |
| 200 | 2.12 | 0.28 | 0.53 | 0.46 | × | ✓ | 100.9 | 97.8 | 199.9 | 0.120 Ne | 2.9 | PISN | × | ⋯ |
| | | | | | $\omega = 0.3$ | | | | | | | | | |
| 100 | 2.68 | 0.25 | 0.45 | 0.0 | × | × | 50.7 | 44.6 | 99.9 | 0.739 O | 3.1 | PPISN | BH | 39.6 |
| 150 | 2.31 | 0.24 | 0.51 | 0.07 | × | × | 79.3 | 70.7 | 149.9 | 0.042 Ne | 3.2 | PISN | × | ⋯ |
| 200 | 2.13 | 0.28 | 0.52 | 0.40 | × | × | 108.3 | 98.1 | 199.9 | 0.002 C | 3.3 | PISN | × | ⋯ |
| | | | | | $\omega = 0.4$ | | | | | | | | | |
| 100 | 2.70 | 0.26 | 0.44 | 0.18 | × | ✓ | 50.3 | 43.7 | 99.9 | 0.775 O | 3.1 | PPISN | BH | 39.4 |
| 150 | 2.34 | 0.24 | 0.50 | 0.36 | × | ✓ | 68.6 | 67.1 | 149.9 | 0.883 O | 2.7 | PISN | × | ⋯ |
| 200 | 2.15 | 0.23 | 0.51 | 0.37 | × | ✓ | 89.6 | 89.6 | 199.9 | 0.042 Ne | 2.7 | PISN | × | ⋯ |
| | | | | | | | | | | | | DBH[e] | BH | 180.0 |
| | | | | | $\omega = 0.5$ | | | | | | | | | |
| 100 | 2.75 | 0.26 | 0.43 | 0.26 | × | ✓ | 49.7 | 43.4 | 99.9 | 0.708 O | 3.1 | PPISN | BH | 39.0 |
| 150 | 2.36 | 0.25 | 0.48 | 0.84 | × | ✓ | 64.3 | 61.6 | 149.9 | 0.882 O | 2.7 | PPISN[d] | BH | 35.8 |
| | | | | | | | | | | | | PISN[c] | × | ⋯ |
| | | | | | | | | | | | | DBH[e] | BH | 135.0 |
| 200 | 2.17 | 0.24 | 0.50 | 0.45 | × | ✓ | 85.0 | 83.8 | 199.9 | 0.047 Ne | 2.7 | PISN | × | ⋯ |
| | | | | | | | | | | | | DBH[e] | BH | 180.0 |
| | | | | | $\dot{M}_{max}$ $\omega = 0.0$ | | | | | | | | | |
| 100 | 2.59 | 0.26 | 0.17 | 0.51 | × | ✓ | 37.1 | 34.3 | 95.0 | 0.412 O | 2.8 | fCCSN[a] | BH | 85.5 |
| | | | | | | | | | | | | PPISN[b] | BH | 30.9 |
| 150 | 2.33 | 0.23 | 0.32 | 0.36 | ✓ | ✓ | 77.5 | 72.6 | 147.7 | 0.019 Ne | 3.1 | PISN | × | ⋯ |
| 200 | 2.15 | 0.22 | 0.30 | 0.35 | ✓ | ✓ | 102.9 | 95.1 | 197.6 | 0.001 C | 3.2 | PISN | × | ⋯ |
| | | | | | $\omega = 0.2$ | | | | | | | | | |
| 100 | 2.66 | 0.25 | 0.46 | 0.004 | × | × | 50.2 | 44.9 | 99.2 | 0.699 O | 3.1 | PPISN | BH | 39.2 |
| 150 | 2.31 | 0.24 | 0.51 | 0.18 | × | × | 78.6 | 71.4 | 147.3 | 0.062 Ne | 3.2 | PISN | × | ⋯ |
| 200 | 2.13 | 0.24 | 0.52 | 0.70 | × | ✓ | 48.7 | 48.7 | 192.0 | 0.834 O | 2.5 | PPISN | BH | 46.7 |
| | | | | | | | | | | | | DBH[e] | BH | 172.8 |
| | | | | | $\omega = 0.3$ | | | | | | | | | |
| 100 | 2.68 | 0.26 | 0.45 | 0.00 | × | × | 50.7 | 44.7 | 99.2 | 0.701 O | 3.1 | PPISN | BH | 39.5 |
| 150 | 2.33 | 0.24 | 0.51 | 0.11 | × | × | 78.6 | 70.0 | 147.6 | 0.024 Ne | 3.2 | PISN | × | ⋯ |
| 200 | 2.14 | 0.24 | 0.53 | 0.51 | × | ✓ | 40.6 | 40.1 | 193.2 | 0.443 O | 2.3 | fCCSN[a] | BH | 173.9 |
| | | | | | | | | | | | | PPISN[b] | BH | 41.8 |
| | | | | | | | | | | | | DBH[e] | BH | 173.9 |
| | | | | | $\omega = 0.4$ | | | | | | | | | |
| 100 | 2.71 | 0.26 | 0.45 | 0.08 | × | × | 50.7 | 44.2 | 99.0 | 0.818 O | 3.1 | PPISN | BH | 39.5 |
| 150 | 2.35 | 0.24 | 0.51 | 0.22 | × | × | 78.8 | 69.6 | 146.9 | 0.035 Ne | 3.1 | PISN | × | ⋯ |
| 200 | 2.16 | 0.24 | 0.53 | 0.81 | × | ✓ | 41.7 | 41.5 | 192.2 | 0.285 O | 2.3 | fCCSN[a] | BH | 173.0 |
| | | | | | | | | | | | | PPISN[b] | BH | 42.4 |
| | | | | | | | | | | | | DBH[e] | BH | 173.0 |
| | | | | | $\omega = 0.5$ | | | | | | | | | |
| 100 | 2.76 | 0.26 | 0.43 | 0.41 | × | ✓ | 49.0 | 42.5 | 97.1 | 0.723 O | 3.1 | PPISN | BH | 38.3 |
| 150 | 2.38 | 0.24 | 0.50 | 0.37 | × | ✓ | 73.1 | 68.7 | 146.0 | 0.884 O | 2.9 | PISN | × | ⋯ |

Table 2
(Continued)

| $M_i$ ($M_\odot$) (1) | $\tau_{MS}$ (Myr) (2) | $\tau_{cHeB}$ (Myr) (3) | $f_{H\,puls}$ (4) | $f_{He\,puls}$ (5) | Blue Loop (6) | DUP (7) | $M_{He}$ ($M_\odot$) (8) | $M_{CO}$ ($M_\odot$) (9) | $M_f$ ($M_\odot$) (10) | $X_{core}$ Onset PI (11) | $L_\nu/L_{rad}$ ($\log_{10}$) (12) | Fate (13) | Remnant (14) | $M_{BH}$ ($M_\odot$) (15) |
|---|---|---|---|---|---|---|---|---|---|---|---|---|---|---|
| 200 | 2.18 | 0.23 | 0.52 | 0.34 | × | ✓ | 52.3 | 51.4 | 194.1 | 0.295 O | 2.3 | PPISN | BH | 48.9 |
| | | | | | | | | | | | | DBH[e] | BH | 174.7 |

**Notes.** Table entries as in Table 1.
[a] Failed CCSN. Following Farmer et al. (2019) we set the lower limit of $M_{He}$ for PPISN at 45 $M_\odot$.
[b] Following Woosley (2017) we set the lower limit of $M_{He}$ for PPISN at 34 $M_\odot$.
[c] Following Woosley (2017) we set the lower limit of $M_{He}$ for PISN at 64 $M_\odot$.
[d] We set the lower limit of $M_{He}$ for PISN at 65.24 $M_\odot$, which is the $M_{He}$ of the T140D model in Woosley (2017).

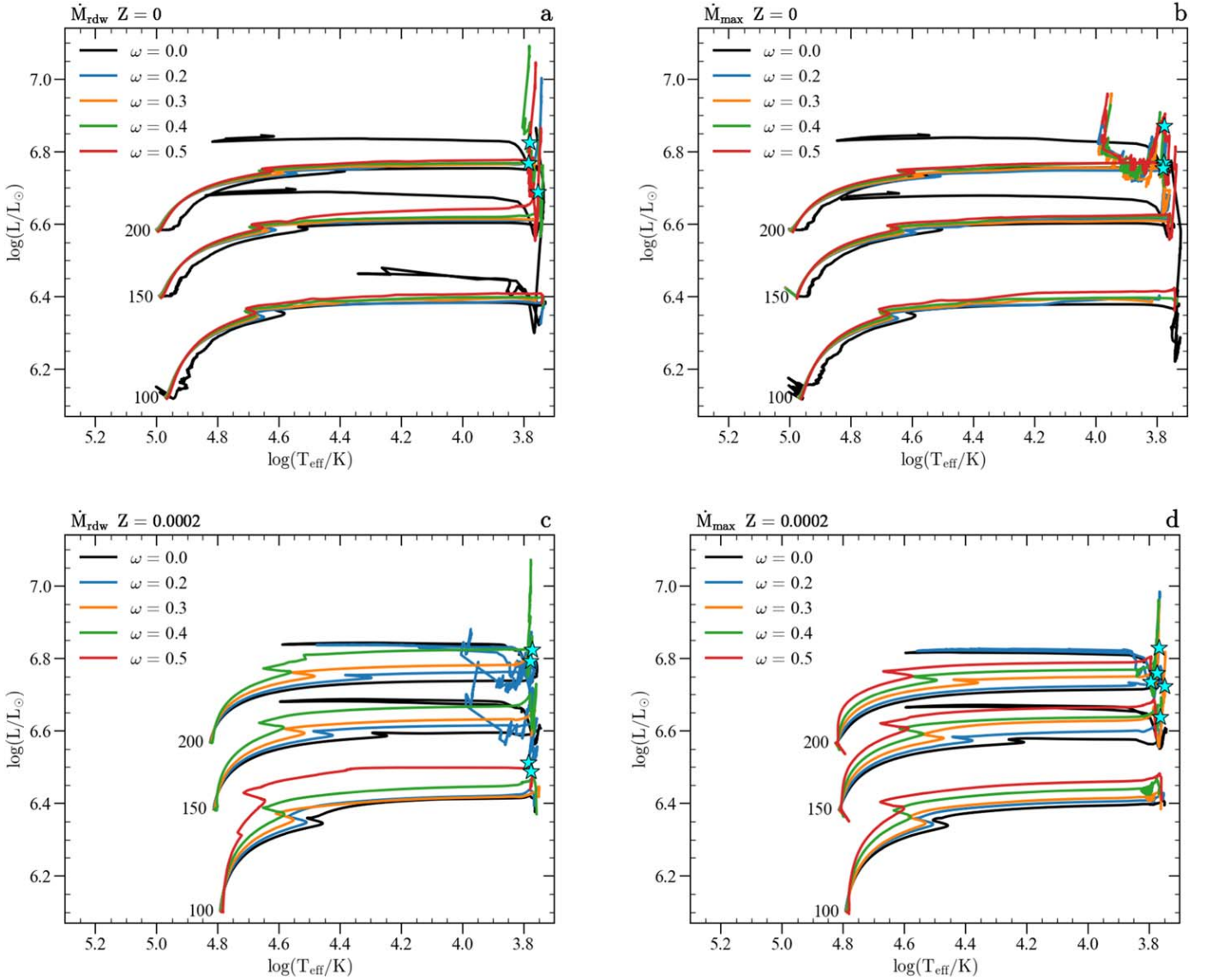

**Figure 1.** H-R diagrams of the 20 sets of tracks computed in this work. Different initial rotation rates are color coded as labeled. Panels (a), (b) and (c), (d) refer to $Z = 0$ and $Z = 0.0002$, respectively. Panels (a), (c): tracks calculated with the standard PARSEC mass-loss prescription for radiation-driven winds. Panels (b), (d): tracks computed considering also the pulsation-driven mass loss. Cyan starred symbols indicate where the stellar models become Wolf–Rayet manqué stars ($X < 0.3$). We plot the value of initial mass (in $M_\odot$) for all tracks.

of the DUP depends on the envelope undershooting parameter, as shown in Costa et al. (2021). According to our definition (see Section 2.1 in Paper I), these 16 stars exhibit a surface chemical composition similar to Wolf–Rayet (WR) stars. Still, they are not hot and spend most of their evolution with a low effective temperature in the red part of the HRD. For this reason, we refer to this kind of star as WR-manqué (WRm; note that the mass loss recipe adopted for these stars is that by de Jager et al. 1988). This drastic change in the surface chemical composition occurs too late to have a great impact on the models' effective temperature evolution. If the hydrogen surface abundance were to decrease below ∼20% during MS (due to high rotation mixing), then the models would have followed the so-called chemically homogeneous evolution. In this case, they would have evolved toward higher effective temperatures, completely avoiding the red-supergiant phase (Woosley & Heger 2006; Yoon et al. 2012). We do not find any occurrence of chemically homogeneous evolution in our models. Among these 16 tracks, six stellar models reach a total amount of hydrogen between 0.26% and 0.1% of their total mass. Section 3.3 examines the impact of dredge-up and rotation on surface abundances and chemical ejecta, especially for these extreme WRm stellar models.

### 3.2. Internal Structure

Figure 2 shows the Kippenhahn diagrams of four different stellar models. They have the same initial mass $M_i = 200\,M_\odot$, but different metallicities, rotational velocities, and mass loss prescriptions. At the beginning of the MS, all these stars have a convective core equal to almost 90% of their total mass. This is because these models have a very large initial mass, and rotation increases the extent of their convective cores. For example, comparing the models in panels (a) and (c), we can notice that the star in this latter panel has a slightly bigger convective core due to its higher initial rotational velocity. Then, the model in panel (a) is evolving almost at constant mass. The reason is that for this model, we adopted the mass-loss recipe $\dot{M}_{\rm rdw}$ with $Z = 0$. On the other hand, for models in panels (b), (c), and (d) the total mass of the star decreases during the evolution. This is most evident for the stars in panels (b) and (d) during the first part of the MS, when the pulsation-driven mass loss is operating.

After the depletion of hydrogen in the stellar core and the start of the cHeB phase, these four models experience DUP episodes. In Figure 2 there are different degrees of envelope penetration, which imply different internal mixing. DUP episodes affect the structure of these models by reducing the mass of the helium core, which can be crucial for the final fate of the stars (see Section 3.4). They also impact the chemical composition of these four models. Depending on the efficiency of the DUP, these stars can remain classical red supergiants, or they can become WRm stars. Despite the DUP, the model in panel (a) retains enough hydrogen at the surface such that it does not become a WRm star, while the opposite occurs for models in panels (b), (c), and (d) (see also Section 3.3). The red vertical line shows when X < 0.3 at the surface, and we see that this does not occur for the model in panel (a). Moreover, the ($M_i = 200\,M_\odot$, $Z = 0$, $\dot{M}_{\rm max}$, $\omega = 0.2$) model in panel (b) becomes an extreme WRm star, with only 0.11% of hydrogen out of its total final mass. Therefore, this star could be considered a pure helium star with a very small carbon–oxygen core (green line in Figure 2(b)) due to the very deep penetration of the stellar envelope.

The surface abundance evolution and wind ejecta masses of all extreme WRm models, along with other selected tracks, are discussed in the Section 3.3.

### 3.3. Surface Chemical Abundances and Mass of the Ejecta

In this section, we consider the surface abundances and the wind ejecta mass of some selected models. Both of these are affected by rotation-induced mixing, dredge-up episodes, and mass-loss history.

Figure 3 presents the surface abundance evolution of H, He, C, N, O, and $Z_{\rm eff} = 1 - X - Y$. Each panel shows the results for the same model computed with two different mass-loss prescriptions, $\dot{M}_{\rm rdw}$ (dotted line) and $\dot{M}_{\rm max}$ (solid line). In Figure 3 there are all six extreme WRm models.

During MS, we see the effect of rotation-induced mixing, especially in the four $Z = 0$ models (panels (a), (b), (c), and (d)). That is, for example, the gradual increase in nitrogen surface abundance, along with a slower enhancement in carbon and oxygen. Approximately ∼$10^5$ yr before the end of computations, all six models start to experience DUP episodes. During cHeB, the hydrogen surface abundance lowers below 0.3 (vertical red line in each panel), while the effective metallicity increases very steeply. At the end of computations, all extreme WRm models in Figure 3 have $Z_{\rm eff} > 0.6$, reaching ∼0.77 in panel (d) for the highest initial rotational velocity. The interesting thing to notice is that at $Z = 0$, the mass-loss prescription accounting also for pulsation-driven mass loss favors the metal enrichment during the evolution, while at $Z = 0.0002$ the extreme WRm models are computed with $\dot{M}_{\rm rdw}$. In panels (c), (d), and (f), both models can be considered WRm stars (two red vertical lines), but only one in each plot is almost completely hydrogen depleted. In these three cases, the extreme WRm models are those that meet the condition $X < 0.3$ earlier on during their evolution. At the end of calculations, these six extreme WRm stars have a surface composition mainly of helium, nitrogen, and oxygen.

Figure 4 shows the wind ejecta mass for He, C, N, O, Ne, and Mg, while tables of wind ejecta for all rotating models can be found on Zenodo.[7] In the six panels of Figure 4 there are all the models presented in Figure 3 along with stellar tracks computed with their same mass and metallicity, but different initial rotational velocity and mass-loss prescription. In panels (b), (c), and (e) are the extreme WRm models discussed in Sections 3.1 and 3.3. In these six panels, we can see the effect of different initial rotational velocities on wind ejecta masses, as well as the impact of the DUP episodes. Instead, panels (a)–(b), (c)–(d), and (e)–(f) show the incidence of the two mass-loss prescriptions $\dot{M}_{\rm rdw}$ and $\dot{M}_{\rm max}$. The general trend is that, also accounting for pulsation-driven mass loss, the ejecta mass of the considered elements increases. This is expected since the definition of the two mass-loss recipes was adopted. Considering the initial metallicity of the models, at higher $Z_i$ we have higher ejecta masses due to higher mass-loss rates. However, the most impactful process for example in panels (a), (b), (c), and (e) is the occurrence of DUP episodes.

The extreme WRm models eject much more metals compared to the others, and this is because of a deeper penetration of the stellar envelope (DUP) coupled to the mass loss history of the model. In panel (a) there are no extreme WRm models, but still, we see major differences between stars computed with $\omega = 0.2$, 0.4, and 0.5 with respect to those with $\omega = 0.0$, and 0.3. The reason for this is the metal enrichment of the models due to DUP episodes (see also Figure 3(a)). Finally, we can say that the differences in the ejecta are due to differences in the mass loss history, except when dealing with models with a very high metal enrichment due to DUP mixing episodes. In these latter cases, the relative ejecta mass of the elements considered is much higher, meaning that mass loss alone can not explain these differences and the main driving process is the mixing due to DUP episodes.

---

[7] doi: 10.5281/zenodo.10225140

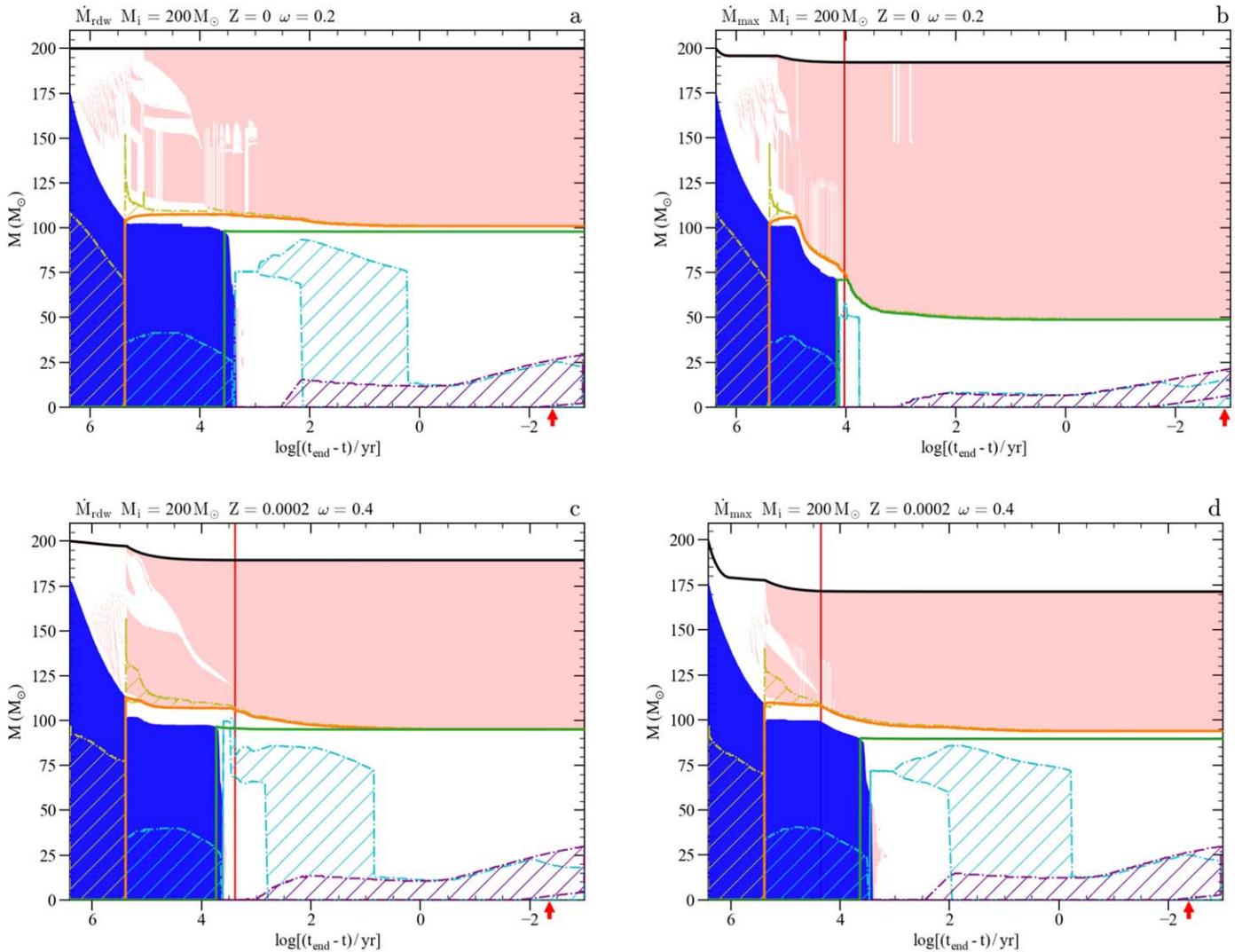

**Figure 2.** Kippenhahn diagrams of four selected models. The horizontal axis shows the age of the models as a logarithm of time (in yr) until the end of computations. The blue regions correspond to the star's convective core; the pink area represents the convective envelope, semiconvective zones at the boundary of the helium convective core, and convective shells. The yellow, cyan, and purple hatch regions show the hydrogen, helium, and carbon-burning core/shells, respectively. The continuous black line indicates the total mass of the star, the orange one represents the helium core, and the green one corresponds to the carbon–oxygen core. The red arrow marks the time when the star enters the unstable region with $\langle \Gamma_1 \rangle = 4/3 + 0.01$; while the red vertical line shows when the star becomes a WRm with $X < 0.3$. Panels (a), (c): models computed with the standard PARSEC mass-loss prescription for radiation-driven winds. Panels (b), (d): models that account also for pulsation-driven winds.

### 3.4. Final Fate

Figure 5 shows the final helium core mass ($M_{He}$) at the end of calculations for all tracks. We use $M_{He}$ as a proxy for the final fate of the star, following Woosley (2017), Farmer et al. (2019). $M_i$ and $M_{He}$ are positively related; however, both panels present stellar models that do not follow this trend. This is caused by DUP episodes that reduce the helium core of the models and in some cases even affect their final fate. The helium core mass is defined by the chemical composition of the envelope, and for this reason, for the 16 WRm stars (see Section 3.1) we consider two extreme possible cases for the final fate. In the first one, as usual, we use the $M_{He}$ core to determine the final fate. In the second case, we use the total final mass of the star at the end of the computations ($M_f$) to derive the final fate. The combined effects of rotation and DUP episodes affect the stellar tracks in different ways. In panel (a), for example, we can see that rotation increases the helium core mass with respect to the nonrotating case for $M_i = 100\,M_\odot$ (see also Table 2). At the same time, DUP episodes and higher mass loss rates prevent a steady growth of the helium core with an increasing initial rotation rate.

Stars with $M_i = 150, 200\,M_\odot$ have He cores that are massive enough to cause a PISN, which disrupts the whole star. There are some exceptions, though. In panel (a), seven tracks do not follow the main pattern. Four of these models were computed with $\dot{M}_{max}$, and only one has $M_i = 150\,M_\odot$. This latter is computed with $\dot{M}_{rdw}$, $\omega = 0.5$ and has three possible fates. If we consider the strict definition of its helium core, this latter has a mass of $64.3\,M_\odot$. With this core, the model is inside the

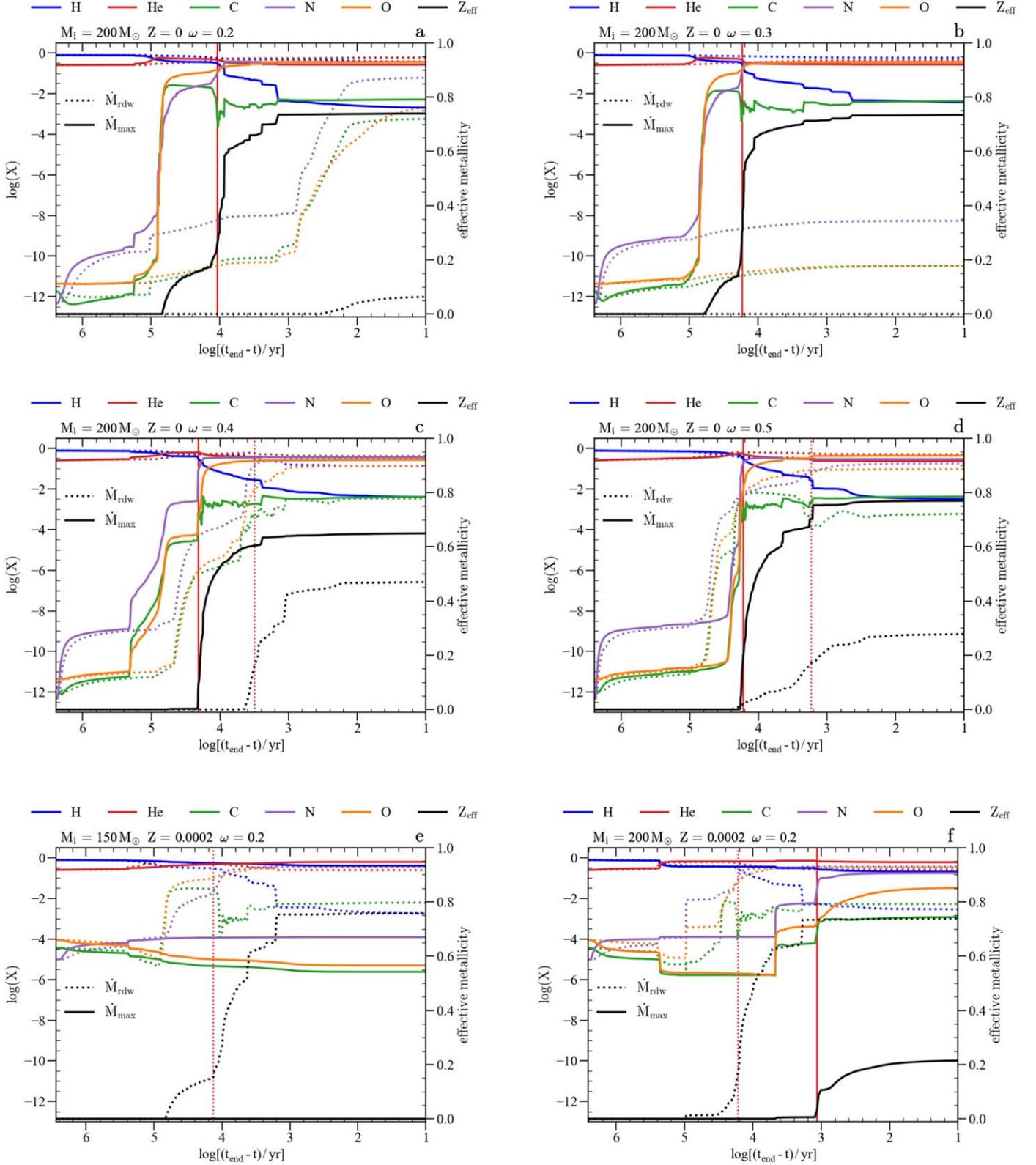

**Figure 3.** Surface chemical abundances' evolution of six selected pairs of models, from the ZAMS to 10 yr before the end of computations. In each panel, the abundances of hydrogen, helium, carbon, nitrogen, and oxygen are color coded as shown in the legend. The black line represents the effective metallicity, $Z_{\rm eff} = 1 - X - Y$. The red vertical line marks when $X < 0.3$. The results are presented for two different mass-loss prescriptions. The horizontal axis is as in Figure 2.

uncertainty strip around the PI—PPI boundary. The lower limit of this strip is set to 63.91 $M_\odot$, while the upper one is 65.24 $M_\odot$. They are the He core masses of the T135D and T140D models from Woosley (2017), respectively. These two models are computed with $Z = Z_\odot/10$ and $\dot{M} = 0$, simulating the evolution of a zero-metallicity star. Even though these two models have a He core mass very close, and in one case exceeding the 64 $M_\odot$ limit, Woosley (2017) found that they produced a stellar-mass

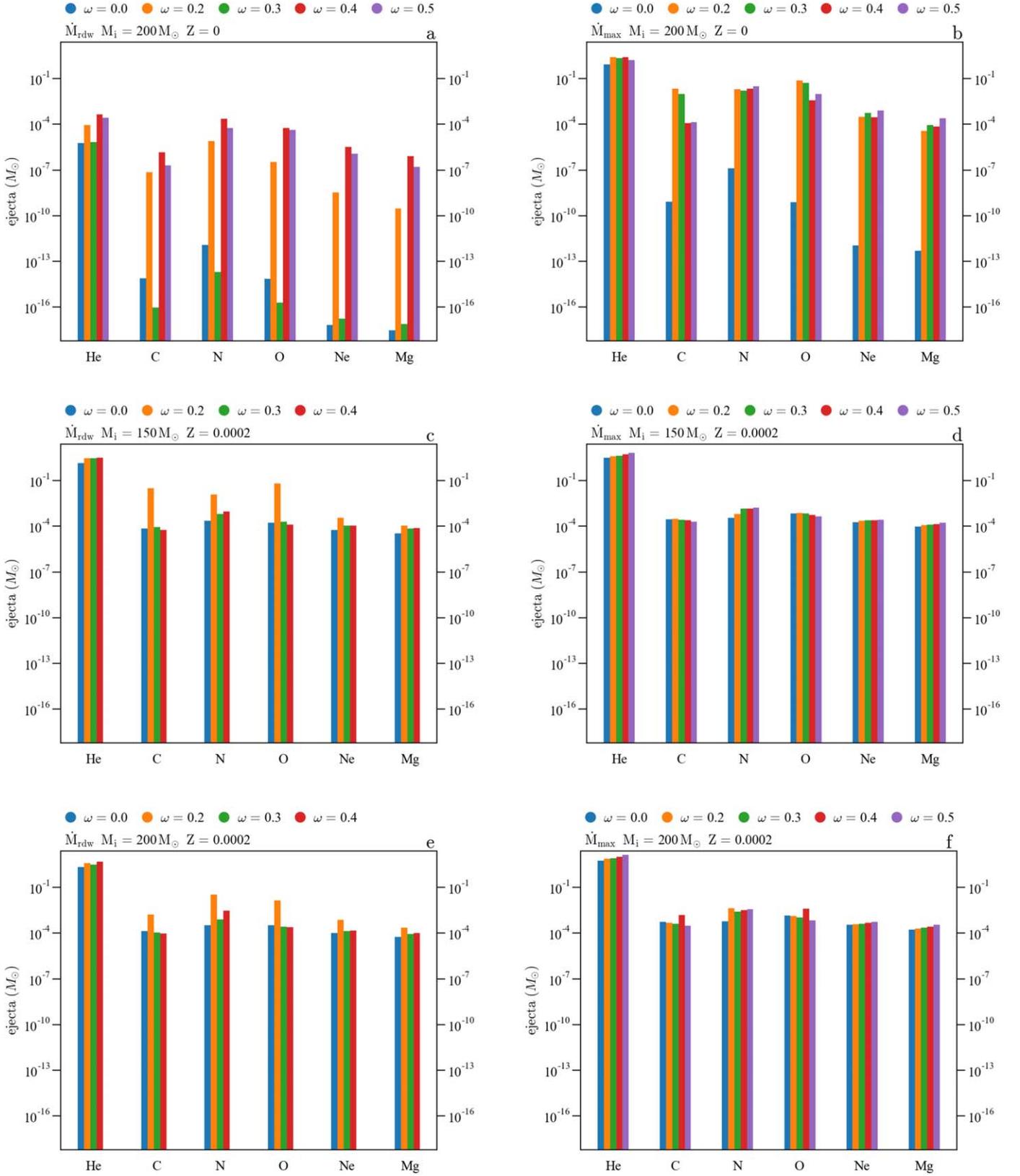

**Figure 4.** Wind ejecta mass of models with initial mass $M_i/M_\odot = 150, 200$. In each panel, there are the ejecta masses of helium, carbon, nitrogen, oxygen, neon, and magnesium for two mass-loss recipes, $\dot{M}_{rdw}$ and $\dot{M}_{max}$. Panels (a), (b): models calculated with $Z = 0$. Panels (c), (d), (e), (f): models calculated with $Z = 0.0002$.

BH. For this reason, we indicate that the stellar model with He core mass between these two limits could have two possible final fates: a PISN with no remnant or a PPISN leaving a stellar-mass BH. On the other hand, the low amount of hydrogen of the $(M_i = 150\, M_\odot, Z = 0, \dot{M}_{rdw}, \omega = 0.5)$ star could imply that the helium core corresponds to the total final mass of the star, $M_{He} = 150\,M_\odot$. In this case, the final fate of the model should be the direct collapse to a BH (DBH). This latter scenario also applies to the ($M_i = 200\,M_\odot$, $Z = 0$, $\dot{M}_{rdw}$, $\omega = 0.4$) and the ($M_i = 200\,M_\odot$, $Z = 0$, $\dot{M}_{rdw}$, $\omega = 0.5$) models if we consider $M_{He} = M_f$ (see Tables 1 and 2 for the different final fates considered for each model).

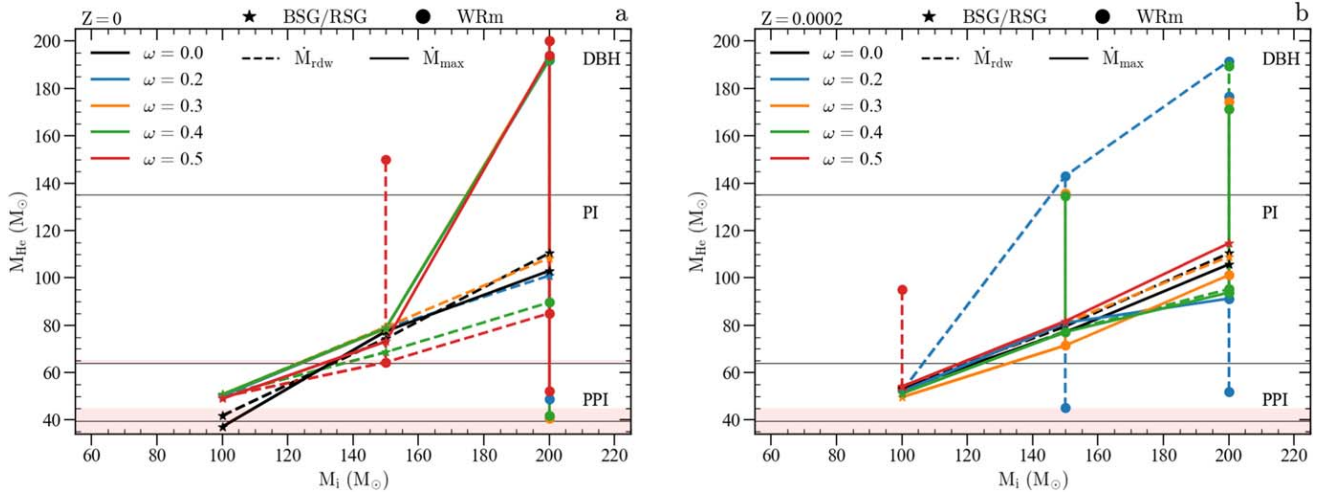

**Figure 5.** Each panel shows the helium core mass, $M_{He}$, as a function of the initial mass for all models of five different initial rotation rates with a fixed initial metallicity. Helium core mass is shown at the end of the calculations. Horizontal lines limit the pulsational pair-instability (PPI), pair-instability (PI) explosion, and direct collapse to a BH (DBH) regime from bottom to top (Woosley 2017; Farmer et al. 2019, 2020). The lower red strip indicates the uncertainty range of the lower limit for PPI. Lower and upper boundaries are 34 $M_\odot$ (Woosley 2017) and 45 $M_\odot$ (Farmer et al. 2019), respectively. The black line is the average. Panel (a): similar uncertainty strip between PPI and PI with boundaries 63.91 $M_\odot$ and 65.24 $M_\odot$. These values are the helium core masses of the T135D and T140D models from Woosley (2017), respectively (see text for more details).

The other tracks in panel (a) with $M_i > 100\,M_\odot$ that do not become PISN are the models ($M_i = 200\,M_\odot$, $Z = 0$, $\dot{M}_{max}$) with $\omega = 0.2, 0.3, 0.4$ and $0.5$. These four models are extreme WRm stars due to their almost H-free envelopes (see Sections 3.1 and 3.3). By considering their helium core masses, these models should undergo PPISN or failed core-collapse supernova (fCCSN, see Figure 5 and Tables 1 and 2). On the other hand, in the case where $M_{He} = M_f$, their helium cores are above the 135 $M_\odot$ upper threshold for PISN; thus, these stars should directly collapse, forming a massive BH.

In panel (b) there are nine WR-manqué stars, five of which are computed with $\dot{M}_{max}$. Both the $M_i = 150\,M_\odot$ and $M_i = 200\,M_\odot$ models with $\dot{M}_{rdw}$ and $\omega = 0.2$ have $M_{He} > 135\,M_\odot$, in the case of $M_{He} = M_f$. Therefore, these two models *could* undergo DBH. Besides these two, seven stars in this panel have a double final fate. The ($M_i = 150\,M_\odot$, $Z = 0.0002$, $\dot{M}_{max}$, $\omega = 0.3, 0.4$), ($M_i = 200\,M_\odot$, $Z = 0.0002$, $\dot{M}_{max}$, $\omega = 0.2, 0.3, 0.4$) and ($M_i = 200\,M_\odot$, $Z = 0.0002$, $\dot{M}_{rdw}$, $\omega = 0.4$) models can either collapse directly to a BH or get totally disrupted in a PISN, depending on the helium core mass considered. On the other hand, the ($M_i = 100\,M_\odot$, $Z = 0.0002$, $\dot{M}_{rdw}$, $\omega = 0.5$) star is in the PPISN or PISN regime in the case $M_{He} < 64\,M_\odot$ or $M_{He} = M_f$, respectively (see also Table 1).

In both panels of Figure 5, stars with $M_i = 100\,M_\odot$ are in the pulsational pair-instability regime, but outside the uncertainty region. The only exceptions are the two models computed with $\omega = 0.0$ (see discussion in Paper I) and ($M_i = 100\,M_\odot$, $Z = 0.0002$, $\dot{M}_{rdw}$, $\omega = 0.5$) already discussed above.

As in Paper I, we use the fit formula proposed by Mapelli et al. (2020, see Appendix D for details) to compute the BH mass in the PPISN scenario, while also accounting for mass loss due to neutrino emission (10% of the baryonic mass of the proto-compact object, see Fryer et al. (2012), Rahman et al. (2022), and references therein). In DBH cases, we take the final mass of the star as a first approximation for $M_{BH}$, always accounting for mass loss due to neutrino emission. In Tables 1 and 2, we give two (or even three) possible outcomes and the corresponding BH masses for all models with an uncertain fate based on the boundaries in Figure 5. Notice that for the ($M_i = 150\,M_\odot$, $Z = 0.0002$, $\dot{M}_{max}$, $\omega = 0.4$) model, we assume a 1% error on the 135 $M_\odot$ threshold for the upper limit for PISN in the fit formula by Mapelli et al. (2020). It is worth mentioning that these limits are based on nonrotating models, as long as the fit formula from Mapelli et al. (2020). There are different studies on the pair-instability limits and the effects of rotation, e.g., Glatzel et al. (1985), Woosley (2017), Marchant & Moriya (2020), Woosley & Heger (2021). For example, Woosley & Heger (2021) mention that high rotation rates could shift the lower limit for pair-instability from 64 $M_\odot$ to $\sim$70 $M_\odot$. For the case where the He core mass is close to the 64 $M_\odot$ limit ($M_i = 150\,M_\odot$, $Z = 0$, $\dot{M}_{rdw}$, $\omega = 0.5$), we can not calculate the BH mass with the fit formula from Mapelli et al. (2020). For this reason, we use a linear interpolation between $M_{He}$ and $M_{BH}$ of the models T135D and T140D. In this way, we can give an estimate of the BH mass for the case of PPISN (see also Table 2).

Tables 1 and 2 and Figure 6 show the results obtained with different mass-loss prescriptions and initial rotation rates. In Figure 6, whenever there are multiple symbols for the same stellar model, it indicates that for that particular star, there is more than one possible outcome. For $Z = 0$, the most massive BHs reach $\sim$180 $M_\odot$, while for $Z = 0.0002$ they reach $\sim$172 $M_\odot$. The complex interplay between DUP and rotation sets the final mass of the BHs we expect from our models. It is worth noticing that rotation favors the occurrence of DUP episodes, shifting, in some cases, the mass of the possible remnant from zero to hundreds of solar masses. Within Section 3.5, the remnants' mass will be rediscussed and adjusted according to results from the analysis on possible jet-driven events (see also Tables 3 and 4).

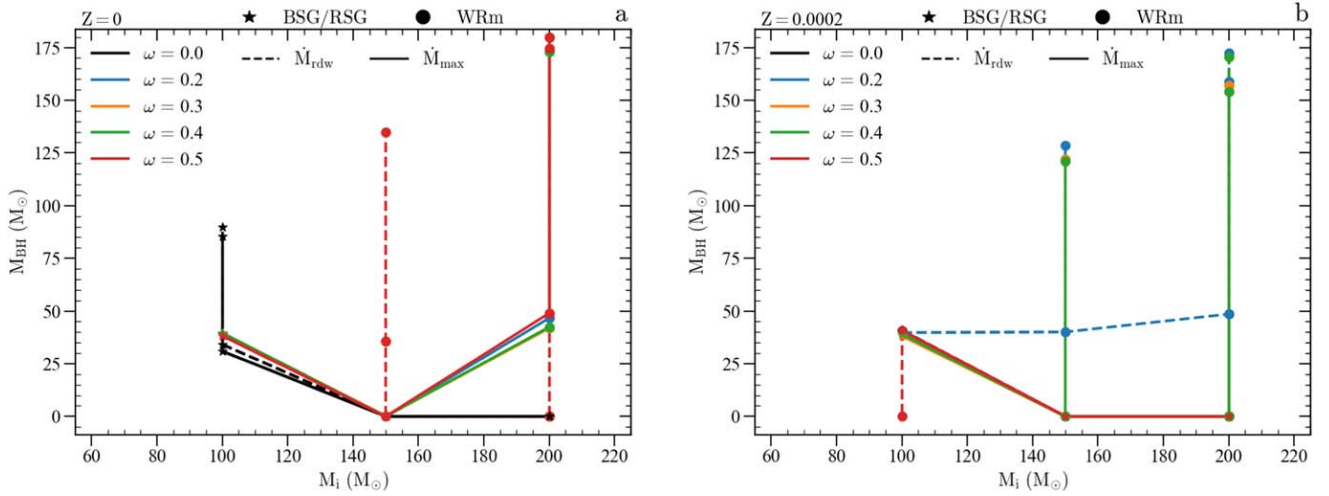

**Figure 6.** Expected remnant mass for all stellar models presented in this work as a function of $M_i$. The extra symbols indicate the predicted BH mass for those stars with multiple possible fates, see also Tables 1 and 2. Panel (a): models calculated with $Z = 0$. Panel (b): models calculated with $Z = 0.0002$.

### 3.5. Progenitors of Jet-driven Events

Within the collapsar scenario (Woosley 1993), the most important characteristics for being a gamma-ray burst (GRB) progenitor are a massive core to produce a BH, the lack of an extended hydrogen envelope to facilitate the jet outward propagation and high core specific angular momentum for the formation of an accretion disk. There are, however, different studies that suggest the possibility of jet propagation even through the very massive envelope of Population III stars (e.g., Ohkubo et al. 2009; Suwa & Ioka 2011; Toma et al. 2011; Nagakura et al. 2012; Wei & Liu 2022). Moreover, there are also multiple criteria for the minimum core specific angular momentum and the mass coordinate to consider when evaluating this latter (e.g., Woosley & Heger 2006; Yoon et al. 2006; Meynet & Maeder 2007; Yoon et al. 2012; Aguilera-Dena et al. 2018; Aloy & Obergaulinger 2021; Obergaulinger & Aloy 2022).

First of all, we have to consider all the shells that have a sufficient specific angular momentum to form a disk instead of falling directly into the newly formed BH (see Appendix B for a different approach in the calculations of the shell inertial moment). Figure 7 shows four example models that summarize all possible cases within our current work. This choice is based on different angular momentum configurations within the progenitors and also the different final fates for the jet-driven events. In each panel, we consider the two limiting cases for the minimum specific angular momentum needed to support mass at the last stable orbit (LSO) of a Schwarzschild and a maximally rotating Kerr BH. Then, the general case considers a BH with mass and angular momentum within the specific mass coordinate in the stellar model (see Bardeen et al. 1972, for the exact expressions). We can see that the specific angular momentum profile is very different between the four models, and therefore the expression $j > j_{crit}$ where we expect disk formation follows different patterns in these four progenitors. We assume that the inner $3\,M_\odot$ of material would form the BH, and therefore do not contribute to the accretion disk. For this reason, we exclude the inner $3\,M_\odot$ from subsequent calculations throughout this work for all our models (see Appendix A, where we consider the case of $M_{BH} = 5\,M_\odot$ for the only model where this implies a sizeable difference).

Having enough angular momentum and forming a disk are necessary but not sufficient conditions for a successful GRB. On top of that, the lack of an extended envelope is needed to allow the jet outlet. To analyze the jet propagation through the progenitor in further detail, we proceed as in Yoon et al. (2015) by computing the accretion rate from the disk. With the approximation introduced by Woosley & Heger (2012), the accretion rate reads

$$\dot{M} = \frac{2M_r}{t_{\rm ff}} \frac{\rho}{\bar{\rho} - \rho}, \qquad (5)$$

with $t_{\rm ff} = 1/\sqrt{24\pi G \bar{\rho}}$ the freefall timescale and $\bar{\rho} = 3M_r/4\pi r^3$ the mean density within each shell. As in Yoon et al. (2015), in this work such accretion rates should be considered as lower limits. This is because our models do not reach the precollapse stage, and hence the density should increase with respect to the considered one.

Figure 8 shows the accretion rate, the freefall timescale, and the crossing timescale for those models presented in Figure 7. Also in these four panels we highlight where the shells have enough specific angular momentum to support a disk (see Figure 7 and related discussion). Panels (b) and (c) do not show the entire progenitor. This is because in the PPISN cases, we plot only the model up to the mass coordinate calculated with the fit formula by Mapelli et al. (2020). In this way, we consider the model after the loss of the envelope due to pulsational pair-instability (Ekström et al. 2008), assuming that the pair-induced pulsations do not affect the angular momentum of the stellar core (Aguilera-Dena et al. 2018). As a first approximation for the velocity of the ejected material due to PPI, we consider the escape velocity at the mass coordinate given by the fit formula from Mapelli et al. (2020). We find that $v_{\rm esc} \sim 10^4\,{\rm km\,s}^{-1}$ and the timescale between the pulsational pair-instability and the collapse of the star should be of the order of $\sim 100$ yr. We can safely stop considering the ejected material in our calculations since it should be at distances of $\sim 1$ pc when the jet forms, thus reducing the density of the ejected envelope by a factor $(1\,{\rm pc})^3$. Instead, in panels (a) and (d) we show the entire stellar model since we consider the

**Table 3**
Most Relevant Properties of Possible GRB Events for Models Computed with $Z = 0.0002$, $\dot{M}_{\rm rdw}$ and $\dot{M}_{\rm max}$

| $M_i$ ($M_\odot$) (1) | $\tau_{\rm cross}$ (s) (2) | $\tau_{\rm free-fall}$ (s) (3) | $\tau_{\rm accretion}$ (s) (4) | $L_{\rm accretion}$ ($\log_{10}({\rm erg\ s}^{-1})$) (5) | $E_{\rm accretion}$ ($\log_{10}({\rm erg})$) (6) | $E_{\rm binding, envelope}$ ($\log_{10}({\rm erg})$) (7) | Fate (8) | $M_{\rm BH}$ ($M_\odot$) (9) |
|---|---|---|---|---|---|---|---|---|
| colspan=9 | | | | | | | | |
| | | | | $\dot{M}_{\rm rdw}$ $\omega = 0.2$ | | | | |
| 100 | 0.35 | 3.35 | 2.45 | 52.48 | 52.87 | ... | c | 39.8 |
| 150 | 46.8 | $5.09 \cdot 10^3$ | 5.66 | 52.52 | 52.74 | 50.21 | a | 40.2 |
|  | $2.32 \cdot 10^3$ | $9.94 \cdot 10^5$ | 5.66 | 52.52 | 52.74 | 50.21 | a[a] | 40.2 |
| 200 | 34.81 | $2.97 \cdot 10^3$ | 6.21 | 52.09 | 52.89 | 51.08[b] | d | 41.4 |
|  | $2.38 \cdot 10^3$ | $8.89 \cdot 10^5$ | 6.21 | 52.09 | 52.89 | 51.08[b] | d[a] | 41.4 |
| | | | | $\omega = 0.3$ | | | | |
| 100 | 0.35 | 3.34 | 2.49 | 52.47 | 52.86 | ... | c | 39.5 |
| 150 | ... | ... | ... | ... | ... | ... | PISN | ... |
| 200 | ... | ... | ... | ... | ... | ... | PISN | ... |
| | | | | $\omega = 0.4$ | | | | |
| 100 | 0.37 | 3.57 | 2.58 | 52.45 | 52.86 | ... | c | 39.4 |
| 150 | ... | ... | ... | ... | ... | ... | PISN | ... |
| 200 | ... | ... | ... | ... | ... | ... | PISN | ... |
|  | $7.43 \cdot 10^3$ | $4.94 \cdot 10^6$ | 13.23 | 52.10 | 53.22 | 50.93 | d[a] | 85.7 |
| | | | | $\omega = 0.5$ | | | | |
| 100 | 0.39 | 3.86 | 2.97 | 52.40 | 52.88 | ... | c | 40.7 |
|  | ... | ... | ... | ... | ... | ... | PISN[a] | ... |
| | | | | $\dot{M}_{\rm max}$ $\omega = 0.2$ | | | | |
| 100 | 0.35 | 3.31 | 2.46 | 52.48 | 52.87 | ... | c | 39.6 |
| 150 | ... | ... | ... | ... | ... | ... | PISN | ... |
| 200 | ... | ... | ... | ... | ... | ... | PISN | ... |
|  | $7.05 \cdot 10^3$ | $4.73 \cdot 10^6$ | 7.72 | 52.39 | 52.44 | 50.88 | a[a] | 82.1 |
| | | | | $\omega = 0.3$ | | | | |
| 100 | 0.35 | 3.37 | 2.59 | 52.43 | 52.85 | ... | c | 38.2 |
| 150 | ... | ... | ... | ... | ... | ... | PISN | ... |
|  | $6.3 \cdot 10^3$ | $4.56 \cdot 10^6$ | 9.47 | 52.11 | 53.09 | 50.69 | d[a] | 64.3 |
| 200 | ... | ... | ... | ... | ... | ... | PISN | ... |
|  | $6.75 \cdot 10^3$ | $4.46 \cdot 10^6$ | 10.46 | 52.22 | 53.24 | 51.02 | d[a] | 91.0 |
| | | | | $\omega = 0.4$ | | | | |
| 100 | 0.36 | 3.53 | 2.66 | 52.44 | 52.86 | ... | c | 39.1 |
| 150 | ... | ... | ... | ... | ... | ... | PISN | ... |
|  | $5.11 \cdot 10^3$ | $3.35 \cdot 10^6$ | 8.94 | 52.17 | 53.12 | 50.93 | d[a](III) | 69.4 |
| 200 | ... | ... | ... | ... | ... | ... | PISN | ... |
|  | $6.77 \cdot 10^3$ | $4.52 \cdot 10^6$ | 11.66 | 52.14 | 53.21 | 50.94 | d[a] | 84.4 |
| | | | | $\omega = 0.5$ | | | | |
| 100 | 0.39 | 3.8 | 2.83 | 52.43 | 52.88 | ... | c | 40.8 |
| 150 | ... | ... | ... | ... | ... | ... | PISN | ... |
| 200 | ... | ... | ... | ... | ... | ... | PISN | ... |

**Note.** The table entries are as follows: (1) star's initial mass; (2) crossing timescale; (3) freefall timescale; (4) accretion timescale, $\sum_i \tau_{\rm acc,i}$; (5) accretion luminosity, $\sum_i L_{\rm acc,i}$; (6) accretion energy, $\sum_i \tau_{\rm acc,i} \cdot L_{\rm acc,i}$; (7) envelope binding energy; (8) case for the possible fate of the GRB progenitor according to four cases outlined in Figure 8; (9) remnant mass.(II)the bottom of the envelope here is defined as the first point where $Y < 10^{-3}$; (III) assuming an error of 1% on upper limit for PISN in fit formula by Mapelli et al. (2020);
[a] Considering $M_{\rm He} = M_{\rm f}$.

$M_{\rm He} = M_{\rm f}$ case and therefore the stars do not undergo PPI (see Section 3.4 and Tables 1, 2).

The models in Figure 8 represent the four possible cases for a jet-driven event among all our rotating stellar tracks, except those that undergo PISN. For this reason, we assign the letters a, b, c, or d in the final fate column of Tables 3 and 4, according to the structure and final fate of the models following one of these four possible cases. Tables 3 and 4 also summarize different physical properties of possible jet-driven events for all rotating models considered in this work.

The model in Figure 8(a) does not experience pulsational pair-instability and it is within the DBH scenario (taking the

Table 4
Most Relevant Properties of Possible GRB Events for Models Computed with $Z=0$, $\dot{M}_{\rm rdw}$ and $\dot{M}_{\rm max}$

| $M_i$ ($M_\odot$) (1) | $\tau_{\rm cross}$ (s) (2) | $\tau_{\rm free-fall}$ (s) (3) | $\tau_{\rm accretion}$ (s) (4) | $L_{\rm accretion}$ ($\log_{10}$(erg s$^{-1}$)) (5) | $E_{\rm accretion}$ ($\log_{10}$(erg)) (6) | $E_{\rm binding,envelope}$ ($\log_{10}$(erg)) (7) | Fate (8) | $M_{BH}$ ($M_\odot$) (9) |
|---|---|---|---|---|---|---|---|---|
| \multicolumn{9}{c}{$\dot{M}_{\rm rdw}$ $\omega=0.2$} |
| 100 | 0.36 | 3.4 | 0.65 | 52.52 | 52.34 | … | b | 39.5 |
|  |  |  | 1.66 | 52.32 | 52.54 |  |  |  |
| 150 | … | … | … | … | … | … | PISN | … |
| 200 | … | … | … | … | … | … | PISN | … |
| \multicolumn{9}{c}{$\omega=0.3$} |
| 100 | 0.36 | 3.46 | 2.67 | 52.44 | 52.87 | … | c | 39.6 |
| 150 | … | … | … | … | … | … | PISN | … |
| 200 | … | … | … | … | … | … | PISN | … |
| \multicolumn{9}{c}{$\omega=0.4$} |
| 100 | 0.36 | 3.51 | 3.12 | 52.37 | 52.86 | … | c | 39.4 |
| 150 | … | … | … | … | … | … | PISN | … |
| 200 | … | … | … | … | … | … | PISN | … |
|  | $7.39 \cdot 10^3$ | $4.77 \cdot 10^6$ | 11.04 | 52.15 | 53.19 | 50.92 | d[a] | 80.6 |
| \multicolumn{9}{c}{$\omega=0.5$} |
| 100 | 0.36 | 3.47 | 3.01 | 52.38 | 52.86 | … | c | 39.0 |
| 150 | … | … | … | … | … | … | PISN[b] | … |
|  | $6.78 \cdot 10^3$ | $4.85 \cdot 10^6$ | 9.12 | 52.08 | 53.04 | 50.61 | d[a] | 57.9 |
|  | 0.31 | 2.85 | 2.08 | 52.50 | 52.82 | … | c[c] | 35.8 |
| 200 | … | … | … | … | … | … | PISN | … |
|  | $7.73 \cdot 10^3$ | $5.10 \cdot 10^6$ | 9.86 | 52.17 | 53.17 | 50.84 | d[a] | 76.5 |
| \multicolumn{9}{c}{$\dot{M}_{\rm max}$ $\omega=0.2$} |
| 100 | 0.35 | 3.31 | 0.64 | 52.50 | 52.30 | … | b | 39.2 |
|  |  |  | 1.40 | 52.23 | 52.38 |  |  |  |
| 150 | … | … | … | … | … | … | PISN | … |
| 200 | 5.31 | $1.80 \cdot 10^2$ | 0.07 | 52.40 | 51.25 | … | d[a] | 46.7 |
|  | $2.34 \cdot 10^3$ | $8.66 \cdot 10^5$ | 0.07 | 52.40 | 51.25 | 50.43 | d[a] | 58.6 |
| \multicolumn{9}{c}{$\omega=0.3$} |
| 100 | 0.35 | 3.38 | 2.66 | 52.44 | 52.86 | … | c | 39.5 |
| 150 | … | … | … | … | … | … | PISN | … |
| 200 | 54.73 | $6.31 \cdot 10^3$ | 0.77 | 52.75 | 52.36 | 50.7 | a[e] | 36.5 |
|  | $2.97 \cdot 10^3$ | $1.24 \cdot 10^6$ | 0.77 | 52.75 | 52.36 | 50.7 | a[a-d] | 36.5 |
| \multicolumn{9}{c}{$\omega=0.4$} |
| 100 | 0.36 | 3.41 | 2.82 | 52.41 | 52.86 | … | c | 39.5 |
| 150 | … | … | … | … | … | … | PISN | … |
| 200 | 37.66 | $3.57 \cdot 10^3$ | 5.74 | 52.08 | 52.84 | 50.84 | d[e] | 37.5 |
|  | $2.83 \cdot 10^3$ | $1.15 \cdot 10^6$ | 5.74 | 52.08 | 52.84 | 50.84 | d[a-d] | 37.5 |
| \multicolumn{9}{c}{$\omega=0.5$} |
| 100 | 0.36 | 3.52 | 3.14 | 52.35 | 52.85 | … | c | 38.3 |
| 150 | … | … | … | … | … | … | PISN | … |
| 200 | 6.34 | $2.30 \cdot 10^2$ | 6.92 | 52.10 | 52.94 | … | c | 49.0 |
|  | $2.77 \cdot 10^3$ | $1.11 \cdot 10^6$ | 6.92 | 52.10 | 52.94 | 50.9 | d[a] | 47.1 |

**Notes.** Table entries as in Table 3.
[a] Concerning the expected BH mass, the model follows the c case since it should experience PPI.
[b] Following Woosley (2017) we set the lower limit of $M_{\rm He}$ for PISN at 64 $M_\odot$.
[c] We set the lower limit of $M_{\rm He}$ for PISN at 65.24 $M_\odot$, which is the $M_{\rm He}$ of the T140D model in Woosley (2017).
[d] Following Farmer et al. (2019) we set the lower limit of $M_{\rm He}$ for PPISN at 45 $M_\odot$.
[e] Following Woosley (2017) we set the lower limit of $M_{\rm He}$ for PPISN at 34 $M_\odot$.

upper limit $M_{\rm He}=M_{\rm f}$, note in Table 3 that the final fate should not change even considering the lower value for $M_{\rm He}$). Therefore, we plot the entire stellar structure where the final mass of the model is 143 $M_\odot$. Panel (a) shows that there could be three distinct accretion episodes due to the distribution of specific angular momentum within the star from matter at

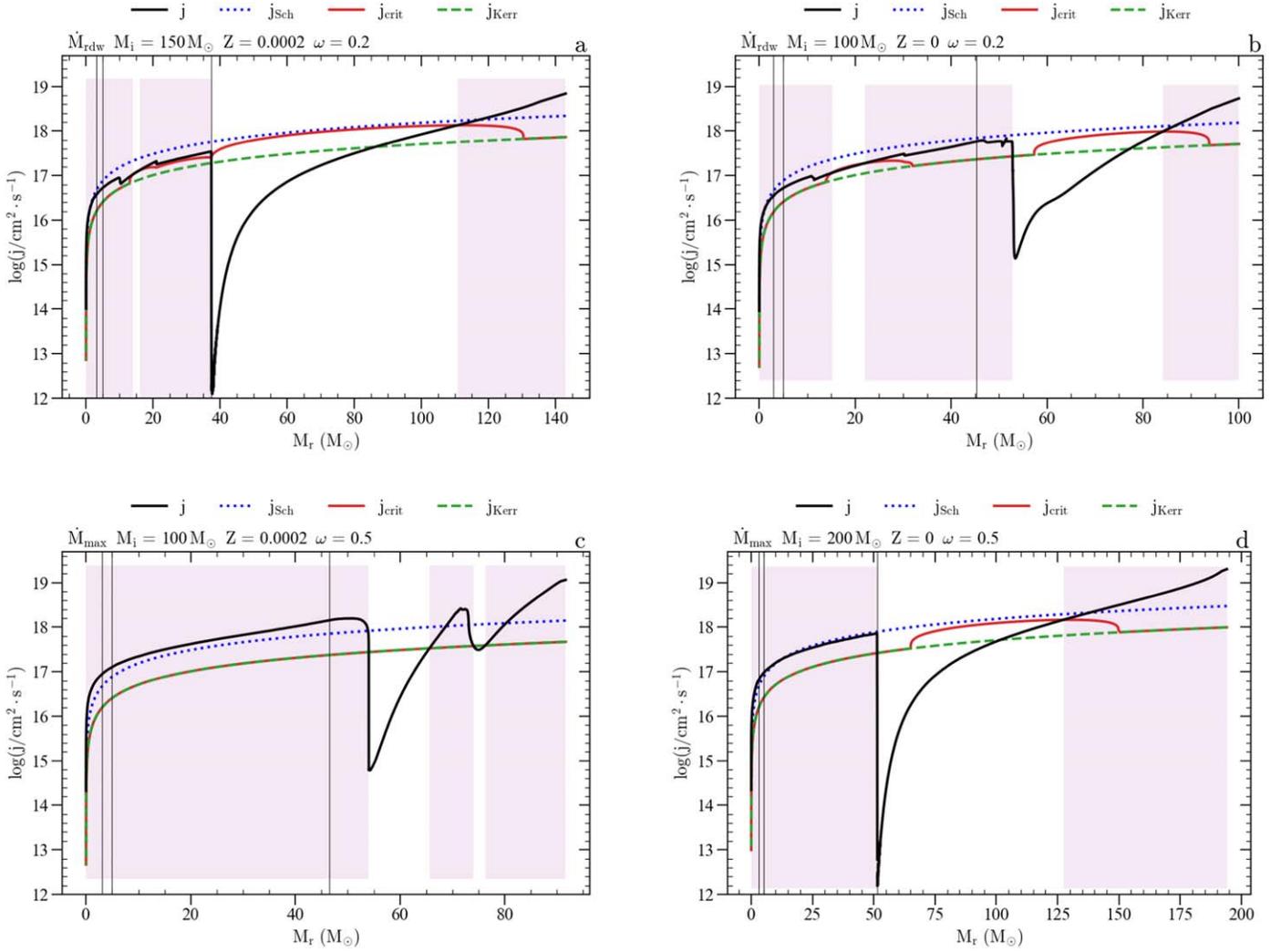

**Figure 7.** Specific angular momentum profile as a function of the mass coordinate. In each panel the black line shows the specific angular momentum for the model considered, the blue line refers to the minimum angular momentum needed to support mass at the LSO of a Schwarzschild BH, while the green line denotes the minimum angular momentum for a maximally rotating Kerr BH. The red line shows the specific angular momentum at the LSO for a BH with mass and angular momentum inside the considered mass coordinate in the computed stellar model. All panels show in lilac the regions within the stars where $j > j_{crit}$. In these regions, we expect the formation of an accretion disk. The two innermost vertical lines in each panel refer to the mass coordinate $3\,M_\odot$ and $5\,M_\odot$, respectively, while the outer one corresponds to the carbon–oxygen core of the model considered.

$3\,M_\odot$–$14\,M_\odot$, $16\,M_\odot$–$38\,M_\odot$ and $111\,M_\odot$–$143\,M_\odot$, respectively.

Combining the first two accretion episodes, the accretion timescale for matter within the stellar core is 5.66 s. However, this is much lower compared to the crossing timescale for jet propagation through the envelope, which is $\sim 10^3$ s (see Table 3). This shows that the jets powered by core accretion are not expected to produce a successful GRB, due to a much difficult jet propagation. Moreover, the crossing timescale is more than two orders of magnitude smaller than the freefall timescale ($\tau_{cross} \ll \tau_{ff}$), and thus the jets would not remain collimated during propagation within the star (Yoon et al. 2015). Following Suwa & Ioka (2011), to calculate the energy of a jet, we assume the following expression for the luminosity:

$$L = \eta\, \dot{M}\, c^2, \qquad (6)$$

where $\dot{M}$ is the mass accretion rate from Equation (5) and the accretion-to-jet conversion efficiency is $\eta = \eta_0\, a^2 = 10^{-3}\, a^2$, where $a$ is the dimensionless spin parameter ($a = J \cdot c / (G \cdot M^2)$) of the central BH with $M_{BH} = 3 M_\odot$ and $J$ the corresponding angular momentum. The first term $\eta_0$ comes from Suwa & Ioka (2011), while the dependence on the BH spin is from Blandford & Znajek (1977). In those models where there are two accretion episodes within the stellar core, to calculate the luminosity of the second jet, we assume $M_{BH} = 3\,M_\odot + M_{acc}$, where $M_{acc}$ is the total mass accreted by the BH before the second accretion episode. Thus, the accretion-to-jet efficiency changes too, because the dimensionless spin parameter is not calculated for a BH of $3\,M_\odot$ as for the other jets. The energy that the jets pump into the stellar envelope is $\sim 3.3 \cdot 10^{52}$ erg. Because these jets can not break out from the progenitor and they are going to spread out instead of remaining collimated, we sum their energy to compare it with the binding energy of the envelope. We define the bottom of the stellar envelope as the first point where $X < 10^{-3}$ (note the only exception in Table 3 to estimate more accurately $M_{BH}$), hence the total envelope binding energy is $\sim 1.6 \cdot 10^{50}$ erg. Because of this, the whole envelope would be ejected by the jets' energy, therefore hindering the third accretion episode. In this case, the final outcome of the model should be a jet-driven supernova (SN).

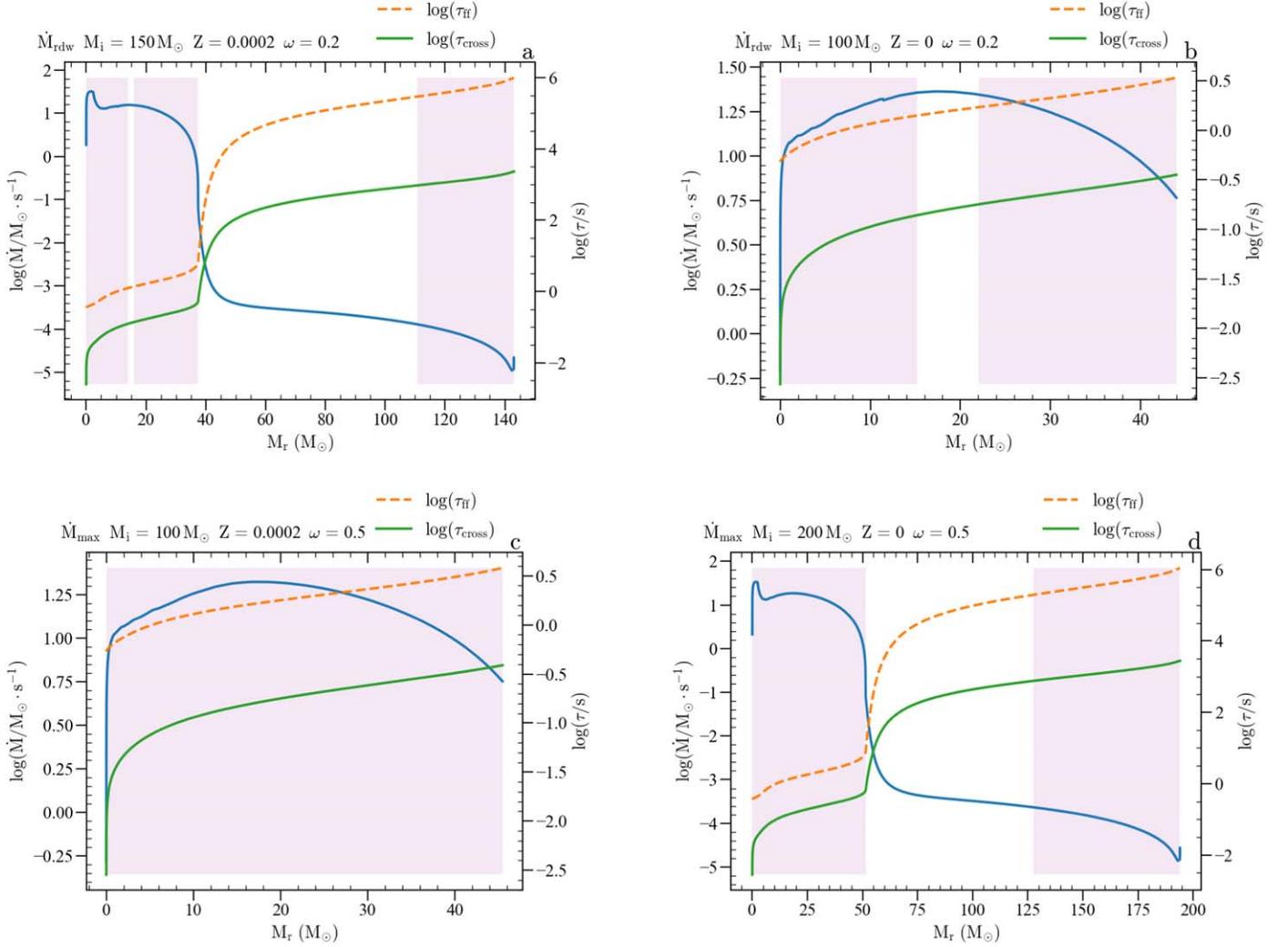

**Figure 8.** Accretion rate, freefall timescale, and crossing timescale as a function of the mass coordinate. Lilac regions show the shells with sufficient specific angular momentum to support a disk, where $j > j_{crit}$. The two timescales refer to the y-axis on the right, while the accretion rate is shown on the left.

In panel (b), the model has enough angular momentum in its inner core to support the formation of an accretion disk. The first accretion episode lasts 0.65 s, while the crossing timescale for the jet to reach the surface of the star is 0.36 s. This means that in this case, there could be a successful GRB event powered by the accretion of matter between $3\,M_\odot$ and $15\,M_\odot$. The crossing timescale is more than four times smaller compared to the accretion timescale for matter between $22\,M_\odot$ and $44\,M_\odot$. This implies that the second accretion episode can not form another jet before the first one breaks out from the star. Here we assume as an upper limit that the first jet does not blow away any mass of the outer stellar core, though this mass could be very small (Zhang et al. 2004). Thus, in this case, we have two accretion episodes that produce successful GRBs. "Case b" corresponds to progenitors of successful GRB events that are powered by two accretion episodes, where in Tables 3 and 4 there are the results for the two jets in the $\tau_{acc}$, $L_{acc}$, and $E_{acc}$ columns.

Case c is very much similar to the latter. The only difference is that the entire model has enough angular momentum to form a disk: there is a single accretion episode that involves the whole star. Even in this case, the accretion timescale is higher concerning the crossing one, 2.83 and 0.39 s, respectively. Therefore, this model should also produce a successful GRB event. Here we assume that the whole mass of the star is going to be accreted, and this gives an upper limit for the accretion timescale and hence the total accretion energy of GRB progenitors that follow the kind of structure presented in this case c.

The model in Figure 8(d) is very similar to that in panel (a) (taking the upper limit $M_{He} = M_f$). As between cases b and c, the most important difference is that in Figure 8(d) there should be only one accretion episode from matter within the stellar core. Even in this case, the accretion timescale is much shorter compared to the crossing timescale (see Table 4). Hence, also in this model, the jet powered by core accretion should produce a jet-driven SN, to which we refer as final fate d in Tables 3 and 4. On the other hand, if we consider the lower value of $M_{He}$ for this specific model, the jet-driven event changes deeply. Because of the lack of an extended envelope, the jet could reach the stellar surface and produce a successful GRB (see Table 4 for the numerical details).

Tables 3 and 4 also present the expected mass of the remnant after the jet-driven event, where also here we account for neutrino emission and take 90%.[8] of the considered final mass for the expected $M_{BH}$ Fryer et al. (2012), Rahman et al. (2022), and references therein. For jet-driven SNe in cases a and d, we consider only the mass of the core because the envelope is ejected by the energy of the jet(s). Instead, in cases b and c, we take into account the total mass of the model after pulsational pair-instability mass loss. All values in the last column in Tables 3 and 4 are to be considered as upper limits for the BH mass. This is because there could be more mass ejected by the jet, even though it might not be that large (Zhang et al. 2004).

Tables 3 and 4 show the luminosity due to accretion-powered jets for all rotating models considered (except those that undergo PISN). The two brightest GRB events have $\log(L) \sim 52.5$. The main driver of the jet-powered events is the accretion rate shown in Figure 8, which depends on the density of the models.

Now we consider different values for the accretion-to-jet conversion efficiency parameter. With $\eta_0 = 10^{-2}; 10^{-4}$, the overall results do not change. Of course, we obtain respectively higher and lower luminosities and accretion energies, but the cases highlighted in Tables 3 and 4 remain unchanged. There is only one exception, which is the $(M_i = 200\,M_\odot, Z = 0, \dot{M}_{max}, \omega = 0.2)$ model in the case of the DBH scenario (see Table 4). With $\eta_0 = 10^{-4}$ the accretion energy results lower than the binding energy of the stellar envelope. Thus in this case the expected remnant mass from this progenitor is higher, $78\,M_\odot$. It is worth noting that this latter is almost $20\,M_\odot$ above the value expected when the total stellar envelope is ejected by the energy of the jet. Considering this lower efficiency, we should expect a failed jet-driven SN from this progenitor, which greatly impacts the final mass of the BH.

## 4. Concluding Remarks

In this work, we study the evolution of rotating zero-metallicity and metal-poor very massive stars with initial masses between $100\,M_\odot$ and $200\,M_\odot$. We investigate the resulting effect arising from different initial rotation rates $\omega = 0.0, 0.2, 0.3, 0.4, 0.5$, and pulsation-driven winds (following the prescription of Nakauchi et al. 2020), accounting also for radiation-driven winds throughout the entire HRD. These are the first very massive models that consider both stellar rotation and pulsation-driven mass loss, extending the parameter space covered by the PARSEC evolutionary tracks. We follow the evolution until the occurrence of electron–positron pair-instability after carbon, neon, or oxygen ignition depending on the initial mass, metallicity, rotation, and mass-loss prescription adopted for the models (see Tables 1 and 2). For all models, we checked that rotation and radial pulsations should not influence each other, with the former dominating over the latter (see discussion in Appendix C).

As in Paper I we discuss the final fate of these stars (Section 3.4), but then accounting also for stellar rotation we analyze the possibility of jet-driven phenomena from these rotating progenitors (Section 3.5). We consider the accretion-to-jet efficiency parameter $\eta = \eta_0 a^2$, where $\eta_0 = 10^{-2}, 10^{-3}, 10^{-4}$, while $a$ is the dimensionless spin parameter of the central BH with $M_{BH} = 3\,M_\odot$ (see Appendix A where we explore the $M_{BH} = 5\,M_\odot$ case). Results in Tables 3 and 4 assume the standard value of $\eta_0 = 10^{-3}$ from Suwa & Ioka (2011).

---
[8] We assume this for consistency with the discussion in Section 3.4.

Following the analysis of Yoon et al. (2015), we identify four possible cases among our rotating models with different structures and final fates.

For stars that do not lose any mass due to pair-instability and still form a BH (DBH scenario, cases a and d), the expected outcome is a jet-driven SN. In these cases, the extended stellar envelope prevents the jet breakout, but the energy of this latter is higher than the envelope binding energy, and thus the jet unbinds the outer layers of the star. The progenitors of these successful jet-driven SNe have $M_i = 200\,M_\odot$, except for four models with $M_i = 150\,M_\odot$ (see Tables 3 and 4). Within case d when considering $\eta_0 = 10^{-4}$, there is only one exception to the fact that the accretion energy of the jet is bigger than the envelope binding energy, i.e., $(M_i = 200\,M_\odot, Z = 0, \dot{M}_{max}, \omega = 0.2)$ model when considering $M_{He} = M_f$. This implies an increase of the expected BH mass of almost $20\,M_\odot$.

All models that undergo pulsational pair-instability lose their envelope, hence fostering the propagation of the jet as opposed to the models in Yoon et al. (2015). Case b and c correspond only to models with $M_i = 100\,M_\odot$, except the $(M_i = 150\,M_\odot, Z = 0, \dot{M}_{rdw}, \omega = 0.5)$ model, which has three possible fates (see Table 4). The major difference between the b and c scenarios is that in the former the star does not have enough angular momentum throughout the whole structure, and thus there are two distinct accretion episodes. Instead, in the latter the whole mass of the star is accreted through the disk in one single episode, which increases the accretion timescale. This difference also occurs between models in cases a and d.

A GRB event can be detected by the BAT X-ray detector up to redshift $z \sim 20$ if it has $L \gtrsim 10^{52}\,\mathrm{erg\,s^{-1}\,s}$ (Komissarov & Barkov 2010; Yoon et al. 2015). All successful GRB events listed in Tables 3 and 4 would be luminous enough to be detected. This changes as we consider different values for the parameter $\eta_0$. When adopting a lower value for efficiency, all the models get below the observability threshold above. On the other hand, increasing the accretion-to-jet efficiency boosts the possibility of detecting this kind of event. Also, the afterglow of these GRBs could be of paramount importance. The reason is that with a bigger energy budget, and therefore a radio flux peaked at late times at gigahertz frequencies (Toma et al. 2011), the radio afterglow of GRBs from Population III progenitors should be much brighter than that of Population II/Population I stars (Salvaterra 2015; Burlon et al. 2016). Hence, this could be key for distinguishing GRB events from different progenitor populations. With current instruments like the Australian Square Kilometer Array Pathfinder telescope and James Webb Space Telescope with both Near-InfraRed Camera and Near-InfraRed Spectrograph, the observation of the afterglow from Population III GRB events should be within reach (Macpherson & Coward 2017).

Jet-driven events also have a deep impact on the expected $M_{BH}$ for some of these progenitors. Comparing the results for the expected BH mass in Tables 1 and 2 with those in Tables 3 and 4, there are 15 models with a different remnant mass. These stars are all within the a and d cases (see discussion in Section 3.5), while for models following cases b and c, the expected mass of the BH does not change. The differences are due to jet-driven SNe that unbind the stellar envelope during the accretion-powered jet's propagation. Hence, the BH mass can be reduced by more than $130\,M_\odot$ in the most extreme cases.

Out of 15 models with a reduced $M_{BH}$, 13 stars have an expected remnant mass that falls within the pair-instability black hole mass gap ($40–65\,M_\odot$ to $120\,M_\odot$, see also Croon et al. (2020), Farmer et al. (2020), Sakstein et al. (2020), Costa et al. (2021), Farrell et al. (2021), Tanikawa et al. (2021), Vink et al. (2021), Farag et al. (2022), for different formation scenarios). Several stellar models can produce BHs with a mass close to one of the primary and the secondary BHs in the gravitational wave event GW190521 (Abbott et al. 2020), which are $85^{+21}_{-14}\,M_\odot$ and $66^{+17}_{-18}\,M_\odot$, respectively. Therefore, rotating primordial very massive stars could provide a new pathway for the formation of BHs within the pair-instability black hole mass gap.

### Acknowledgments

We are grateful to Riccardo Ciolfi and Martin Obergaulinger for their very precious comments and helpful discussions. This research is supported by the Italian Ministerial grant PRIN 2022, "Radiative opacities for astrophysical applications," no. 2022NEXMP8, CUP C53D23001220006. P.M., G.V., M.T., and L.G. acknowledge support from Padova University through the research project PRD 2021. G.C. acknowledges support from the Agence Nationale de la Recherche grant POPSYCLE number ANR-19-CE31-0022. L.G. acknowledges partial funding by an INAF Theory grant 2022. A.B. acknowledges support by PRIN MIUR 2017 prot. 20173ML3WW 001.

*Software:* PARSEC (Bressan et al. 2012; Costa et al. 2019a, 2019b, 2021), OPAL (Iglesias & Rogers 1996), ÆSOPUS (Marigo & Aringer 2009; Marigo et al. 2022), FREEEOS (A.W Irwin; http://freeeos.sourceforge.net ).

## Appendix A
## An Alternative Case for Central Black Hole Mass

Here we investigate the case study where we assume $M_{\rm BH}=5\,M_\odot$ instead of $3\,M_\odot$, while keeping $\eta = 10^{-3}\,a^2$. The different mass of the central BH has two major effects on the calculations for the jet-driven episode. First, there are two solar masses less for the BH to accrete, which impacts the total rate of mass accretion and the accretion timescale. Taking out some of the material with a high accretion rate reduces $\dot{M}$ while increasing $\tau_{\rm acc}$. The second effect is that in the luminosity calculations, we have to take into account the dimensionless spin parameter derived for the inner $5\,M_\odot$. This in principle has a different value compared to $a(M_{\rm BH}=3\,M_\odot)$.

For all our rotating models, considering $M_{\rm BH}=5\,M_\odot$ has very little effect on the results shown in Tables 3 and 4. At most there is a difference of $\sim 0.5$ s in $\tau_{\rm acc}$ and $\sim 0.01$ in $\log(L_{\rm acc})$. The only exception is the ($M_i = 200\,M_\odot$, $Z=0$, $\dot{M}_{\rm max}$, $\omega = 0.2$) model when we consider $M_{\rm He} = M_{\rm f}$, whose timescales and accretion rates are shown in Figure 9. We see that there are no shells with $j > j_{\rm crit}$ within the core outside $5\,M_\odot$. The model has sufficient angular momentum only in the very outer envelope, but the accretion rate there is much lower compared to the inner core. Since all the inner layers of the star would fall into the BH right away, the jet formed by the accretion of the outer layers would not have to pierce any stellar envelope and therefore in this case the crossing timescale is zero.

In this particular case, the model could be a progenitor of a successful GRB but much fainter than those presented above due to the lower accretion rate. The luminosity is $L \sim 9.3 \cdot 10^{45}$ erg s$^{-1}$, and this transient could last for more than two weeks because of the very long accretion timescale, $\sim 1.4 \cdot 10^6$ s. This different GRB event also impacts the expected BH mass from this progenitor. Contrary to cases a and d (see Section 3.5), the envelope is not ejected, and therefore we have to consider the 90% of the total mass of the model. This increases the mass of the black hole to $M_{\rm BH} = 172.8\,M_\odot$.

On the other hand, in the PPISN case, if we assume $M_{\rm BH}=5\,M_\odot$ then in this model there are no shells with enough angular momentum able to form an accretion disk. Hence, the whole stellar core should collapse into a BH.

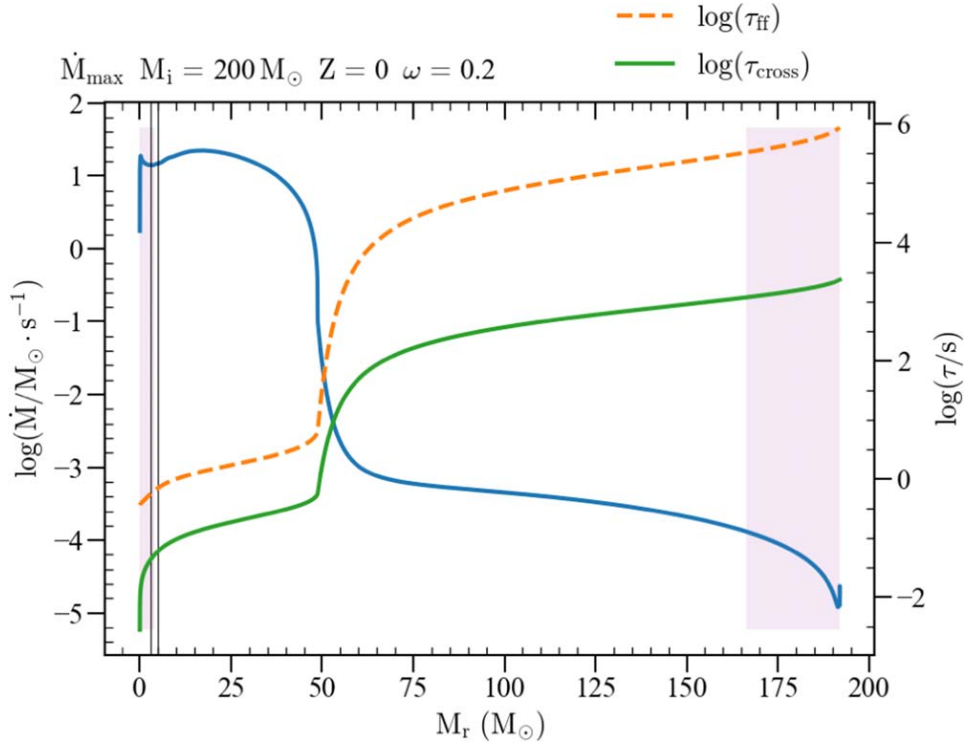

**Figure 9.** Same as in Figure 8. The two black vertical lines are at $3\,M_\odot$ and $5\,M_\odot$, respectively.

## Appendix B
## GRB Analysis with Improved Moment of Inertia

Here we show a different approach in the moment of inertia calculations for the shells in two different models (see in Figure 10). In Section 3.5, we evaluate the specific angular momentum profile, averaging the inertia moment within each shell. We improve our analysis by splitting the integral for the moment of inertia into a polar region and an equatorial region as follows:

$$I_{\rm pol} = 2 \cdot 2\pi \rho \frac{R_1^5 - R_2^5}{5} \int_0^{\pi/4} \sin^3(\theta) d\theta, \quad (B1)$$

$$I_{\rm eq} = 2 \cdot 2\pi \rho \frac{R_1^5 - R_2^5}{5} \int_{\pi/4}^{\pi/2} \sin^3(\theta) d\theta, \quad (B2)$$

with $\rho = \frac{3}{4\pi}(M_1 - M_2)/(R_1^3 - R_2^3)$ where $M_1$, $M_2$, $R_1$, and $R_2$ are the limiting masses and radii for the considered shell. The specific angular momentum profile is $j_{\rm eq(pol)} = I_{\rm eq(pol)} \cdot \omega / M$, where $\omega$ is the angular velocity and $M$ the mass of the shell. In this way, in Figure 10 there are two different profiles for the specific angular momentum in the polar (purple) and equatorial (cyan) regions. As expected, we can see that $j_{\rm eq} > j_{\rm pol}$ since the angular momentum is more concentrated in the equatorial region.

To assess the possibility of a successful GRB, we have to consider two different aspects. First, whether the equatorial region has enough angular momentum to form a disk and therefore to launch a jet. Second, whether the material in the polar region collapses directly into the BH, thus favoring the jet propagation.

For every mass coordinate, there could be three possible scenarios, depending on the relations between $j_{\rm eq}$, $j_{\rm pol}$ and $j_{\rm crit}$. The best scenario for a successful GRB would be when $j_{\rm eq} > j_{\rm crit}$ and $j_{\rm pol} < j_{\rm crit}$. In this situation, we would have an accretion disk that powers the jet and no stellar matter that hampers its propagation through the poles. In the second scenario, both the equatorial and polar regions have enough angular momentum to prevent direct accretion onto the BH. Here the jet would not break out and therefore we could not have a successful GRB. Finally, we could have the case where both $j_{\rm eq}$ and $j_{\rm pol}$ are smaller than $j_{\rm crit}$. In this case, the jet could not be launched since there would be no accretion disk.

We also checked that with this improved analysis, all our rotating models are still described by the four possible cases presented in Section 3.5. The results in Tables 3 and 4 are still valid except for the models ($M_i = 150\,M_\odot$, $Z = 0.0002$, $\dot{M}_{\rm rdw}$, $\omega = 0.2$), ($M_i = 200\,M_\odot$, $Z = 0.0002$, $\dot{M}_{\rm rdw}$, $\omega = 0.2$), ($M_i = 200\,M_\odot$, $Z = 0$, $\dot{M}_{\rm max}$, $\omega = 0.2$), ($M_i = 200\,M_\odot$, $Z = 0$, $\dot{M}_{\rm max}$, $\omega = 0.3$), and ($M_i = 200\,M_\odot$, $Z = 0$, $\dot{M}_{\rm max}$, $\omega = 0.4$) that in the PPISN scenario should follow the b/c cases discussed in Section 3.5.

In Figure 10, we present two models already discussed in Section 3.5 to show the differences with respect to the previous analysis (they are the models in Figure 7(c) and (d), respectively). In panel (a), the star undergoes PPI, so we have to consider only the stellar core. Here we see that the equatorial region has always enough angular momentum to form a disk. Instead, the material in the polar region should collapse directly into the BH. This is the most favorable scenario discussed above. On the other hand, in panel (b) we have to consider the whole star since this model should not experience PPI when we take $M_{\rm He} = M_{\rm f}$. Within the CO core, the situation is very similar to that in panel (a). However, the presence of the stellar envelope should prevent the jet from breaking out.

Figure 11 shows the corresponding accretion rates, crossing timescales, and freefall timescales. There are two main differences with respect to the models presented in Figure 8. Since we split each shell into a polar and an equatorial region, only half of the total shell mass can be accreted through the disk and power the jet. For this reason, the mass accretion rate is a factor of $1/2$ lower. This implies a difference of $\sim 0.3$ dex in $\log(L_{\rm acc})$. The other difference is in the crossing timescale. All models present a smaller $\tau_{\rm cross}$ for the jet powered by the inner core material. The difference is not that relevant because it does not change the outcome of the jet-driven event (e.g., $\Delta \tau_{\rm cross} \sim 0.14$ s for the first jet in the model $M_i = 100\,M_\odot$, $Z = 0$, $\dot{M}_{\rm rdw}$, $\omega = 0.2$). For models like the one in panel (a), $\tau_{\rm cross}$ is set to zero. The reason is that the polar region is devoid of matter, and thus the jet should freely propagate outward. Similarly, the second possible jet in models following Figure 7(b) should have $\tau_{\rm cross} = 0$.

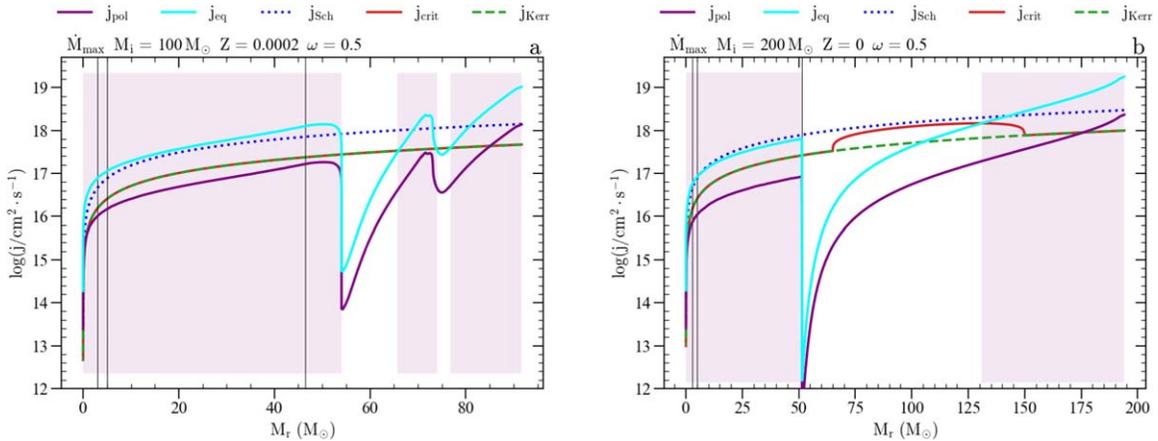

**Figure 10.** The two limiting cases in blue and green, the red curve, and the three vertical lines are as in Figure 7. In each panel, the purple line shows the specific angular momentum for the polar region, while the cyan line displays it for the equatorial region. We highlight in lilac where $j_{\rm eq} > j_{\rm crit}$.

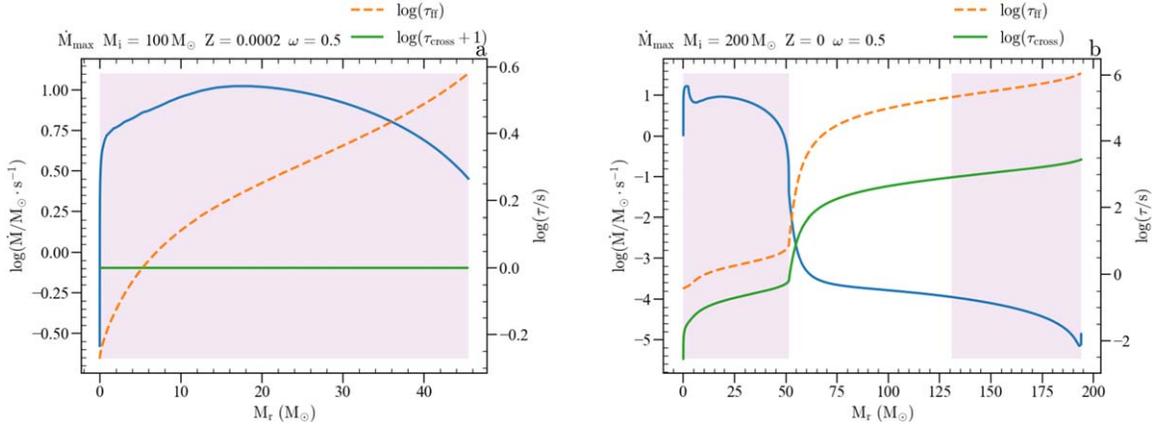

**Figure 11.** Same as in Figure 8, but the lilac regions refer to $j_{eq} > j_{crit}$.

## Appendix C
## Interplay Between Pulsation-driven Mass Loss and Rotation

Currently, no description of the interplay between rotation and stellar oscillations is included in PARSEC. However, it can easily be shown that such effects are negligible under the conditions we are interested in. Indeed, we only consider radial oscillations, which are unaffected by the centrifugal deformation (e.g., Saio 1981; Anderson et al. 2016). The magnitude of the remaining effects scales approximately as $(P/P_{rot})^2$, the squared ratio between the pulsation and rotation periods (Anderson et al. 2016). To assess the importance of such effects, the pulsation period can be approximated by the dynamical timescale $\tau_{dyn} \sim \sqrt{R_*^3/2GM}$ (e.g., Catelan & Smith 2015).

Figure 12 shows the squared ratio between $\tau_{dyn}$ and $P_{rot}$ for all models considered in this work. We see that there is a consistent difference between $\tau_{dyn}$ and $P_{rot}$, with a maximum squared ratio of $\sim 0.038$ in panel (c). The resulting effects should be within at most $\sim 0.4\%$, thus indicating that rotation and radial pulsations should not influence each other.

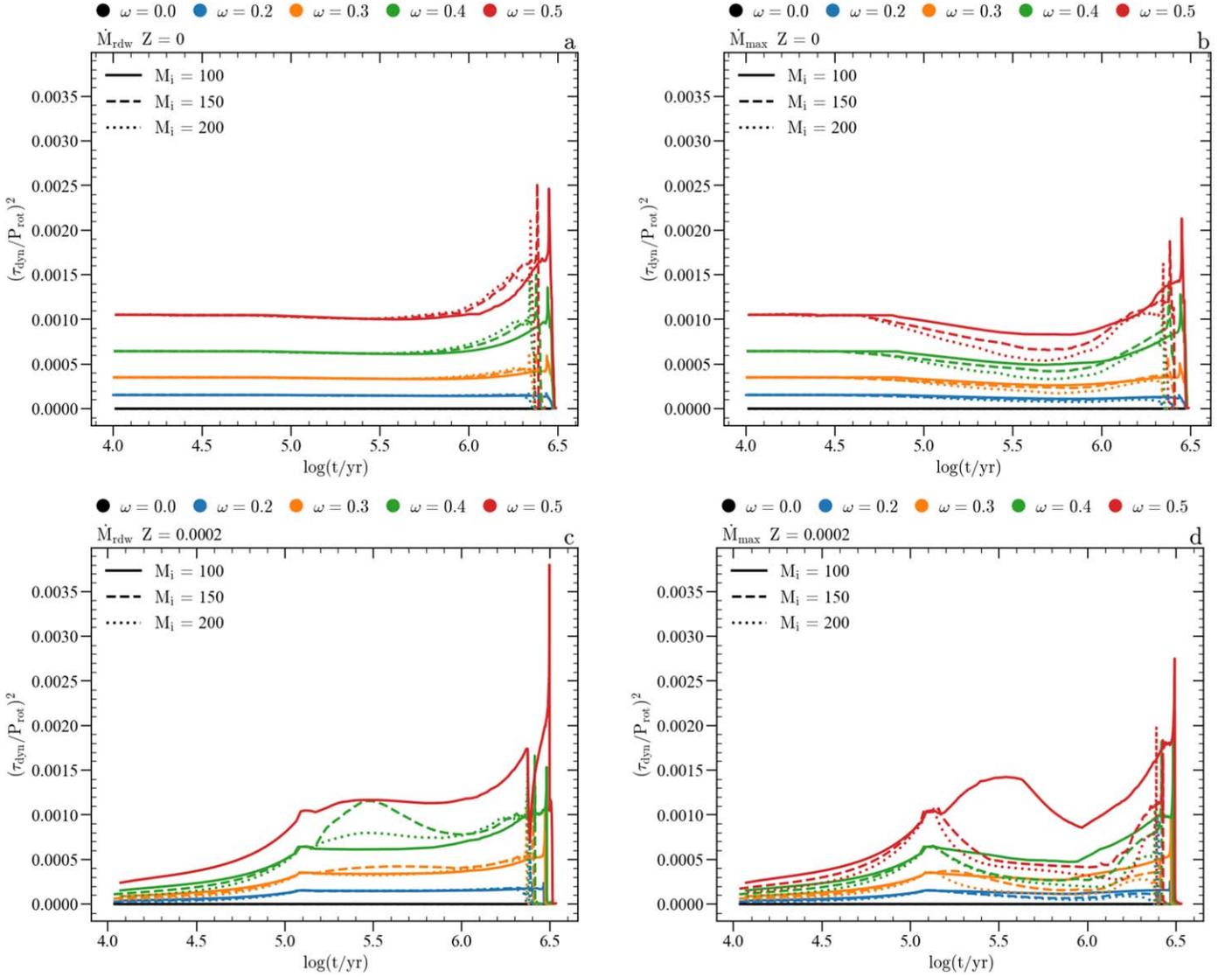

**Figure 12.** Squared ratio between dynamical timescale $\tau_{\rm dyn}$ and rotation period $P_{\rm rot}$ as a function of time ($\log(t/{\rm yr})$) for all stellar models computed in this work. For simplicity and visualization convenience, these panels show $(\tau_{\rm dyn}/P_{rot})^2$ from $t = 10^4$ yr after the beginning of the pre-main sequence until the end of computations. Different colors indicate different initial rotational velocities, while different line-styles refer to the three possible initial masses. The initial metallicity and the mass-loss recipe adopted in the sets are indicated at the top of each panel. The three possible initial masses are in units of $M_\odot$.

## Appendix D
## Fit Formula by Mapelli et al. 2020

We adopt the fitting formula by Mapelli et al. (2020), which relies on models by Woosley (2017), to calculate the expected remnant mass of our models. The general expression for the mass of the remnant is $m_{\rm rem} = \alpha_{\rm P}\, m_{\rm no\ PPI}$, where $m_{\rm no\ PPI}$ is the mass we would obtain without considering PPISN/PISN. In this work, we assume $m_{\rm no\ PPI} = 0.9\, M_{\rm f}$, where $M_{\rm f}$ is the final mass of the model, to account for mass loss due to neutrino emission (see Fryer et al. 2012; Rahman et al. 2022, and references therein).

The fitting formula is a function of the helium core mass $M_{\rm He}$, the final mass $M_{\rm f}$, and the parameters $F$, $K$, and $S$ defined as follows

$$F = \frac{M_{\rm He}}{M_{\rm f}},\ K = 0.67000 F + 0.10000,\ S = 0.52260 F - 0.52974. \quad (D1)$$

Then, the coefficient $\alpha_{\rm P}$ reads

$$\alpha_{\rm P} = \begin{cases} 1 \text{ if } M_{\rm He} \leqslant 32\ M_\odot,\ \forall\ F,\ \forall\ S \\ 0.2(K-1)M_{\rm He} + 0.2(37 - 32K) \text{ if } 32 < M_{\rm He}/M_\odot \leqslant 37,\ F < 0.9,\ \forall\ S \\ K \text{ if } 37 < M_{\rm He}/M_\odot \leqslant 60,\ F < 0.9,\ \forall\ S \\ K(16.0 - 0.25 M_{\rm He}) \text{ if } 60 < M_{\rm He}/M_\odot < 64,\ F < 0.9,\ \forall\ S \\ S(M_{\rm He} - 32) + 1 \text{ if } M_{\rm He} \leqslant 37 M_\odot,\ F \geqslant 0.9,\ \forall\ S \\ 5S + 1 \text{ if } 37 < M_{\rm He}/M_\odot \leqslant 56,\ F \geqslant 0.9,\ 5S + 1 < 0.82916 \\ (-0.1381 F + 0.1309)(M_{\rm He} - 56) + 0.82916 \text{ if } 37 < M_{\rm He}/M_\odot \leqslant 56,\ F \geqslant 0.9,\ 5S + 1 \geqslant 0.82916 \\ -0.103645 M_{\rm He} + 6.63328 \text{ if } 56 < M_{\rm He}/M_\odot < 64,\ F \geqslant 0.9,\ \forall\ S \\ 0 \text{ if } 64 \leqslant M_{\rm He}/M_\odot < 135,\ \forall\ F,\ \forall\ S \\ 1 \text{ if } M_{\rm He} \geqslant 135 M_\odot,\ \forall\ F,\ \forall\ S. \end{cases} \qquad (D2)$$

## ORCID iDs

Guglielmo Volpato https://orcid.org/0000-0002-8691-4940
Paola Marigo https://orcid.org/0000-0002-9137-0773
Guglielmo Costa https://orcid.org/0000-0002-6213-6988
Alessandro Bressan https://orcid.org/0000-0002-7922-8440
Michele Trabucchi https://orcid.org/0000-0002-1429-2388
Léo Girardi https://orcid.org/0000-0002-6301-3269
Francesco Addari https://orcid.org/0000-0002-3867-9966